\documentclass[12pt,english]{article}
\pdfoutput=1

\usepackage[authordate,
backend=biber,
doi=only,
isbn=false,
sorting=nyt,
maxcitenames=3,
minbibnames=7,
maxbibnames=7,
uniquename=false,
sortcites=true]{biblatex-chicago}

\bibliography{bib/references.bib}

\AtEveryBibitem{\clearlist{note}\clearlist{language}\clearlist{issn}} 
\AtEveryBibitem{%
	\ifentrytype{online}{%
		\clearfield{urlyear}
		\clearfield{urlmonth}
		\clearfield{urlday}
		\clearfield{note}
		\clearlist{language}
	}{%
		\clearfield{eprint}%
		\clearfield{urlyear}
		\clearfield{urlmonth}
		\clearfield{urlday}
		\clearfield{note}
		\clearlist{language}
	}
}%

\usepackage{mathptmx}
\usepackage{microtype}
\usepackage{amsmath}
\usepackage{amssymb}
\usepackage{bbm}
\usepackage{bm}
\usepackage[utf8]{inputenc}
\usepackage[T1]{fontenc}
\usepackage{babel}
\usepackage{setspace}
\usepackage{graphicx}
\usepackage{longtable,booktabs,threeparttablex}
\usepackage{array}
\usepackage{multirow}
\usepackage[usenames,dvipsnames,svgnames,table]{xcolor}
\usepackage[export]{adjustbox}[2011/08/13]
\usepackage{enumitem}
\usepackage[text={16cm,24cm}]{geometry}
\usepackage{ragged2e}
\usepackage{csquotes}

\usepackage[hang, flushmargin, bottom, symbol]{footmisc}
\usepackage[colorlinks=true, linkcolor=blue, citecolor=blue, plainpages=false, pdfpagelabels=true, urlcolor=blue]{hyperref}
\usepackage{float}

\usepackage{subcaption}
\usepackage{caption}
\makeatletter
\date{September 8, 2021}

\newtheorem{theorem}{Theorem}
\newtheorem{assumption}{Assumption}
\newtheorem{definition}{Definition}

\newtheorem{proof}{Proof}

\usepackage{footnotebackref}
\newcommand{\PAPERKEYWORDS}{\textbf{Keywords}: Economic Stimulus Act, American Rescue Plan, Consumption inequality, Propensity to consume, Welfare inequality}
\newcommand{\PAPERJEL}{\textbf{JEL}: I38, E21, C6}

\newcommand{\PAPERTITLE}{Optimal allocations to heterogeneous agents with an application to stimulus checks}

\newcommand{\AUTHORNYGAARD}{Vegard M. Nygaard}
\newcommand{\AUTHORNYGAARDURL}{https://sites.google.com/site/vegardmokleivnygaard/}
\newcommand{\AUTHORNYGAARDINFO}{\href{\AUTHORNYGAARDURL}{\AUTHORNYGAARD}: Department of Economics, University of Houston, Houston, Texas, USA (email: vmnygaard@uh.edu)}

\newcommand{\AUTHORSORENSEN}{Bent E. S{\o}rensen}
\newcommand{\AUTHORSORENSENURL}{https://uh.edu/~bsorense/}
\newcommand{\AUTHORSORENSENINFO}{\href{\AUTHORSORENSENURL}{\AUTHORSORENSEN}: Department of Economics, University of Houston, Houston, Texas, USA and Centre for Economic Policy Research, London, UK (email: besorensen@uh.edu)}

\newcommand{\AUTHORWANG}{Fan Wang}
\newcommand{\AUTHORWANGURL}{https://orcid.org/0000-0003-2640-5420}
\newcommand{\AUTHORWANGINFO}{\href{\AUTHORWANGURL}{\AUTHORWANG}: Department of Economics, University of Houston, Houston, Texas, USA (email: fwang26@uh.edu)}

\newcommand{\ACKNOWLEDGMENTS}{
This paper subsumes and replaces two previously circulated and unpublished manuscripts by the authors: \citetitle{wangOptimalAllocationResources2020} \autocite{wangOptimalAllocationResources2020} and \citetitle{nygaardOptimalAllocationCOVID192020a} \autocite{nygaardOptimalAllocationCOVID192020a}.
We thank the editor, B. Ravikumar, two anonymous referees, Samson Alva, Edmund Crawley, Flávio Cunha, Lei Fang, Willa Friedman, Chinhui Juhn, Fatih Karahan, Dirk Krueger, Yunan Li, Yona Rubinstein, Esteban Puentes, Jim Schmitz, Gianluca Violante, Dietz Vollrath, Pei Cheng Yu, as well as seminar participants at conferences and the Board of Governors, Louisiana State University, Oslo Macro Group, University of Bergen, and University of Massachusetts Lowell for helpful comments.}

\newcommand{\PAPERABSTRACT}{
A planner allocates discrete transfers of size $D_g$ to $N$ heterogeneous groups labeled $g$ and has CES preferences over the resulting outcomes, $H_g(D_g)$. We derive a closed-form solution for optimally allocating a fixed budget subject to group-specific inequality constraints under the assumption that increments in the $H_g$ functions are non-increasing. We illustrate our method by studying allocations of ``support checks'' from the U.S. government to households during both the Great Recession and the COVID-19 pandemic. We compare the actual allocations to optimal ones under alternative constraints, assuming the government focused on stimulating aggregate consumption during the 2008--2009 crisis and  focused on welfare during the 2020--2021 crisis.
The inputs for this analysis are obtained from versions of a life-cycle model with heterogeneous households, which predicts household-type-specific consumption and welfare responses to tax rebates and cash transfers.\\
\PAPERJEL}

\newcommand{\PAPERDOIURL}{https://doi.org/10.1016/j.jedc.2022.104352}
\newcommand{\PAPERINFO}{This paper is published as: Nygaard, Vegard M., Bent E. Sørensen, and Fan Wang. ``Optimal Allocations to Heterogeneous Agents with an Application to Stimulus Checks.'' Journal of Economic Dynamics and Control 138 (May 1, 2022): 104352.
\url{\PAPERDOIURL}.
}

\begin{document}

\title{\PAPERTITLE
\thanks{
\PAPERINFO
}}

\author{
\AUTHORNYGAARD{},
\AUTHORSORENSEN{},
and \AUTHORWANG\thanks{
\AUTHORNYGAARDINFO;
\AUTHORSORENSENINFO;
\AUTHORWANGINFO.
\ACKNOWLEDGMENTS
}}

\date{March 31, 2022}

\maketitle

\begin{abstract}
\singlespacing
\PAPERABSTRACT\\
\PAPERKEYWORDS
\end{abstract}
\thispagestyle{empty}
\clearpage

\pagenumbering{arabic}
\setcounter{page}{1}
\renewcommand*{\thefootnote}{\arabic{footnote}}

\section{Introduction}
A planner desires to allocate a number $D_g$ of discrete units of a resource to heterogeneous groups $g=\{1,\dots,N\}$, with resulting outcomes $H_g(D_g)$ assumed to increase with $D_g$ at non-increasing rates. Each group is typically identified by a vector of covariates, but a group can be a single individual. The planner has a constraint on the total number of transfers, faces limits on how many units can be transferred to each group, and evaluates the distribution of outcomes using a CES index over group outcomes. We derive a closed-form solution for optimally allocating a fixed budget in such situations.

More formally, we consider the
following optimization problem
\begin{equation}
\label{eq:discmaxproblem_cons}
\max_{\{\underline{D}_g \le D_g\le\overline{D}_g\}_{g=1}^N}
\left(
\sum_{g=1}^{N}
\beta_g\left(
H_g (D_g)  \right)^{\lambda}
\right)^{\frac{1}{\lambda}},
\end{equation}
subject to an aggregate resource constraint. Here, $\{\underline{D}_g,\overline{D}_g\}$ are group-specific lower and upper bounds on allocations, $\beta_g$ is a weight that allows for certain groups to be prioritized by the planner, and $\lambda\in(-\infty,1]$ parameterizes the strength of the planner's aversion to inequality in outcomes. We show that for given inputs to the algorithm (i.e., calculated increments in $H_g (D_g)$), there exists a closed-form solution to this problem which takes the form of an \emph{optimal allocation queue} that lists the order in which each increment should be allocated until aggregate resources are exhausted. The optimal allocation queue is a sorted list of these increments, where successive values in the queue can involve the same or different groups.

We illustrate our method by applying it to study the optimal allocation of ``stimulus checks'' (or ``support checks'') from the U.S. government to groups of households defined by covariates such as income and civic status. The two applications that we consider are the Economic Stimulus Act of 2008---where households received stimulus checks in the form of tax rebates in an effort to stimulate the economy in the wake of the Great Recession---and the Coronavirus, Aid, Relief, and Economic Security (CARES) Act of 2020/American Rescue Plan (ARP) Act of 2021---where the government issued stimulus checks in the form of cash transfers as part of an effort to insure households against the economic fallout brought about by the COVID-19 pandemic. We compare actual and optimal allocations under alternative household limits on tax rebates and check amounts. Motivated by the actual policies, we assume that the government in 2008 focused on stimulating aggregate consumption,
whereas the government in 2020--2021 focused on the lifetime utility, or welfare, of check recipients.

To apply our framework and solve for the optimal allocations, we need inputs from a consumer model. For the 2008 tax rebates, we need to know the effect of increments in the rebates on each group's consumption, and for the 2021 cash transfers, we need to know the effect of increments in the transfers on each group's lifetime utility. We derive these inputs by formulating a life-cycle consumption-savings model where consumers are ex-ante heterogeneous in marital status, educational attainment, number of children, and rate of time discounting. Consumers face idiosyncratic shocks to income and fertility over their life-cycle. Unemployment probabilities and unemployment insurance are calibrated to the relevant crisis-periods, and the marginal utility of consumption is assumed to be temporarily reduced during the COVID-19 pandemic due to the lockdown. We use the model to predict the impact of each \$100 increment in tax rebates or check amounts for different household types: single or married, having 0--4 children under age 18, of different ages and income-levels.

A key feature of our approach is to break the problem into two steps. First, approximate the allocations by an integer number of increments (in our application, \$100 increments), and use data, regressions, or a model (in our application, the life-cycle model applied to a typical member of a group) to compute each group's gains (as evaluated by the planner) from each allocation increment within the given limits. Second, based on the matrix of gains for each potential increment, sort those gains across both increments and groups in a descending sequence and optimally allocate from the beginning till the budget is exhausted.

While we focus on the optimal allocation of tax rebates and checks, our solution method can be applied in several settings where a government or agency needs to allocate limited resources among heterogeneous recipients. Resources may come in discrete units or in continuous units for which discrete approximations are reasonable (as in our examples), and there may be bounds on how much the recipients can and/or must receive.
Examples are job training programs for displaced workers and nutritional aids for children who are at risk of undernourishment. Moreover, the framework easily allows for different planner objectives: in our examples, maximizing aggregate consumption and maximizing a CES index over the welfare of recipients. The inputs required by the algorithm can be output from structural models, regressions, experiments, or data. The \href{https://fanwangecon.github.io/PrjOptiSNW/}{companion code package} efficiently solves for the optimal allocation queue along with aggregate outcome statistics for any value of planner inequality-aversion for given inputs.

Our algorithm requires non-increasing impacts of increments, which may rule out some applications, for example, consumer models with threshold wealth needed for acquiring a durable good, and it does not allow for direct spillover effects between groups, which precludes it from being used to evaluate the benefits of, for example, vaccinations which protect recipients as well as those around them.\footnote{In the case of consumption stimulus and welfare, some consumers may be at ``thresholds'' where an extra check allows them to buy a car or even a residence. In our applications, we believe that the groups that we consider based on income, marital status, children, and age, are large enough that such threshold effects will average out over the members of the group; although this might not be the case in an application where checks are based on very detailed recipient characteristics.}

\subsection*{Related literature}
Our main contribution, which rests on solving for optimal policies, is conceptually similar to the large literature that applies quantitative macroeconomic models to evaluate alternative policies. Comparing a broad range of potential policies, however, quickly becomes computationally costly, unless our approach of constructing an optimal allocation queue is applied. Our approach trivially allows one to study a large number of potential policies, each of which might differ in detailed group-specific constraints imposed by the planner, without having to recalculate the underlying inputs required by the algorithm.

For the second step of evaluating the queue, our approach is reminiscent of the ``sufficient statistics for welfare analysis'' literature. Unlike this literature, which typically relies on first order conditions for maxima (see \textcite{chettySufficientStatisticsWelfare2009} for a summary), our approach allows for discrete returns to allocations and can handle policies with a large number of group-specific constraints, for which standard Lagrangian methods may be infeasible or computationally costly to apply.

Our approach is also related to the optimal treatment literature. For example, recent papers have considered optimal treatment rules given budget and policy constraints---see for example \textcite{bhattacharyaInferringWelfareMaximizing2012a, kitagawaWhoShouldBe2018a}. This literature often studies convergence to an optimal allocation when allocation rules are based on observations from a finite sample---an issue that we do not address in our paper---and typically studies binary treatment rules. The idea of the allocation queue might be useful in this context, in particular in settings beyond the binary treatment case.

Our life-cycle consumption-savings model with heterogeneous consumers builds on the incomplete markets literature. Recent papers have utilized these models to study the COVID-19 pandemic. \textcite{carrollModelingConsumptionResponse2021} estimate the consumption response to stimulus checks using a buffer-stock model similar to ours where consumers are impatient and credit constrained, but they do not consider how to optimally allocate the checks. We follow their modeling of the impact of COVID-19 on the marginal utility of consumption. Similarly, several papers study the effects of the fiscal response to the pandemic; for example, \textcite{bayerCoronavirusStimulusPackage2020} who use a HANK model and \textcite{faria-e-castroFiscalPolicyPandemic2021} who uses a two-agent DSGE model---see \textcite{falcettoniLiteratureReviewImpact2021} for a more comprehensive review of this literature.

We briefly summarize select papers that study the consumption response to the 2008 and 2020 tax rebates/stimulus checks. For the 2008 tax rebate, \textcite{sahmHouseholdResponse20082010} surveyed consumers and found that about a third of tax rebates were spent after a year, with significant heterogeneity among consumers.\footnote{In our paper, we will refer to this as the average propensity to consume (APC), noting that this APC differs by tax rebate amount, and we will use the term marginal propensity to consume (MPC) to refer to the marginal increase in consumption following a \$100 incremental increase in the tax rebate or check amount.}
\textcite{brodaEconomicStimulusPayments2014}, using combined survey and scanner data, find even larger consumption responses, especially among consumers with low liquid wealth. \textcite{grazianiWorkersSpendingResponse2016} find an average MPC out of the 2008 tax rebates and the 2011 payroll tax cuts of around one-third after a year.

For the COVID-19 stimulus checks, \textcite{bakerIncomeLiquidityConsumption2020a} find that consumers making less than \$1,000 a month spent about 40 percent of the check within a few weeks, while individuals making more than \$5,000 a month spent about 20 percent. \textcite{coibionHowDidConsumers2020} conduct a large-scale survey of U.S. consumers and find that consumers spent about 40 percent of the check. Consistent with our model's predictions, they find that younger, poorer, and larger households spent more. They point out that ``stimulus payments were less effective because they were larger than previous ones. As the size of one-time transfers to households rises, diminishing returns induces individuals to consume smaller fractions of their temporarily higher income''---a key feature of our consumption model. \textcite{kargerHeterogeneityMarginalPropensity2021} compare spending on credit- and debit cards over the two weeks before and after the checks were deposited and find average spending impacts of around 40 percent.

The paper proceeds as follows. Section~\ref{sec:algorithm} presents the allocation problem and its solution. Section~\ref{sec:model} presents the life-cycle model and the calibration during non-crisis periods. Section~\ref{sec:results_2008_policy} applies the algorithm to the 2008--2009 Great Recession and derives the optimal allocation of tax rebates. Section~\ref{sec:results_2021_policy} applies the algorithm to the 2020--2021 pandemic and derives the optimal allocation of cash checks. Section~\ref{sec:conclusion} concludes the paper. The appendix provides the proof for the optimal allocation queue. Further details about the calibration, additional results, and further technical details are provided in the online appendix.

\section{Allocation problem}\label{sec:algorithm}
\subsection{Planner's allocation problem}\label{subsec:planner-problem}
Consider the problem of a planner whose objective is to allocate a discrete amount of resources, $D_g$, among $N$ groups indexed by $g$. Assume the planner has Atkinson-CES preferences \autocite{atkinsonMeasurementInequality1970a} over group-specific outcomes, $H_g(D_g)$, assumed to increase with $D_g$ at non-increasing rates. This section shows that there exists a closed-form solution to the optimal allocation problem characterized by an \emph{optimal allocation queue}. This queue solves the planning problem for any value of planner inequality-aversion, group-specific weights and allocation constraints, and allows for both continuous and non-continuous returns to allocations.

The constraint set is
\begin{align}\label{eq:disconstrainset}
\mathbb{C}^D \equiv \Bigg\{D_g \in \mathbb{N}_0, \underline{D}_g \le D_g \le \overline{D}_g,\sum_{g=1}^ND_g\leq {W}\Bigg\},
 \end{align}
where ${W}$ is an integer equal to the number of increments that fits within the budget.
It is generally not possible to solve this discrete allocation problem by comparing the value under all possible allocations because the size of $\mathbb{C}^D$ grows factorially with $N$ and $\overline{D}_g$, and evaluating all combinatorial possibilities therefore becomes computationally infeasible.\footnote{At the lower-end, where $\overline{D}_g=1$, the number of allocation possibilities is binomial, $\frac{N!}{\left(N-{W}\right)!{W}!}$. At the upper-end, where $\overline{D}_g={W}$, the number of allocation possibilities is multinomial, $\frac{\left({W}+N-1\right)!}{\left(N-1\right)!{W}!}$.}
Let $A_g$ denote  $H_g(0)$ and define $\alpha_{g,l}$ as the increase in $H_g$ due to allocation increment number $l\geq1$; i.e., $\alpha_{g,l} \, = \,  H_g  (l \,\epsilon) -H_g([l-1] \,  \epsilon)$.\footnote{In the stimulus problem in Section~\ref{sec:results_2008_policy}, with consumers' initial consumption at $C_g$, $\alpha_{g,l}=[ C_g+ \Sigma_{s=1}^l MPC_{g,s}]-[C_g+ \Sigma_{s=1}^{l-1} MPC_{g,s} ] = MPC_{g,l}$.} For convenience, we define $\alpha_{g,0}=0$.\footnote{In the welfare problem in Section~\ref{sec:results_2021_policy}, $A_g$ is consumer $g$'s lifetime utility when receiving no checks $(D_g=0)$ and $\alpha_{g,l}$ is the increase in expected lifetime utility when receiving the $l$'th check increment conditional on having already received $l-1$ increments.} We can then express $H_g(D_g)$ as $H_g(D_g)=A_g + \sum_{l=1}^{{D}_g } \alpha_{g,l}$. We define the solution to the discrete optimal allocation problem as follows:
\begin{definition}\label{def:Solution}
Given $N$ groups, the solution to the discrete optimal allocation problem is a set of allocation functions $D_j^*\left({W},\lambda,\left\{\beta_g\right\}_{g=1}^N,\left\{A_g\right\}_{g=1}^N,
\left\{\left\{\alpha_{g,l}\right\}_{l=1}^{\overline{D}_g}\right\}_{g=1}^N,\left\{\underline{D}_g,\overline{D}_g\right\}_{g=1}^N\right) : \mathbb{N}\times(-\infty,1]\times$ $(0,1)^N\times\mathbb{R}^N\times\mathbb{R}_+^{\sum_{g=1}^N \overline{D}_g}\times\mathbb{N}_0^{2N}$ $\rightarrow$ $\left\{\underline{D}_j,\underline{D}_j+1,\dots,\overline{D}_j\right\}$ such that the vector $(D_1^*,...,D_N^*)$ satisfies
\begin{equation}
\label{eq:discmaxproblem}
\max_{\left\{\underline{D}_g\leq D_g\leq\overline{D}_g\right\}_{g=1}^N}
\left(
\sum_{g=1}^{N}\beta_g
\left(
A_g + \sum_{l=0}^{{D}_g }  \alpha_{g,l} \right)^{\lambda}
\right)^{\frac{1}{\lambda}},
\end{equation}
subject to $\sum_{g=1}^N D_g \leq{W}.$
\end{definition}
\noindent As shown in Definition~\ref{def:Solution}, $A_g$ and $\alpha_{g,l}$ reduce the state space of the underlying dynamic programming problem to a dimension-reduced allocation space.

While the discrete allocation problem differs from a continuous allocation problem, the solution to the discrete allocation problem converges to the solution of a corresponding continuous allocation problem as the size of the allocation increments goes to zero as long as the corresponding $H$-functions are continuous on a compact support. Many allocation problems are discrete or can naturally be formulated as a decision on how many discrete units to allocate to each recipient. For example, in our empirical applications we consider each tax rebate or check to be an integer number, each of which represents a \$100 increment in tax rebate or check amount.
\subsection{Solution to the discrete allocation problem}\label{subsec:Solution_to_discrete_allocation_problem}
We impose three restrictions on $\alpha_{g,l}$ and $A_g$:
\begin{assumption}\label{assumption:alpha_A}
For all $g$ and all $l>0$, marginal effects $\alpha_{g,l}$ for the l'th increment of $D_g$ on $H_g$ are: (1) positive, $\alpha_{g,l}>0$; and (2) non-increasing, $\alpha_{g,l}\leq\alpha_{g,l-1}$ for all $l>1$. (3) $A_g>0$ for all $g$.
\end{assumption}
\noindent Assumption~\ref{assumption:alpha_A} imposes no functional form or parametric assumptions on the underlying reduced form or structural model. The first restriction trivially holds in many applications and is typically easily verified.
The second restriction accommodates both constant returns and arbitrary step functions of decreasing returns. The third restriction can be replaced by $A_g+\sum_{l=0}^{\overline{D}_g-1}\alpha_{g,l}>0$, which generally allows for $A_g<0$, if the planner first allocates the resources required for the cumulative effects of the allocations to lead to positive outcomes.

Our solution concept will be in the form of an optimal allocation queue:
\begin{definition}\label{def:Allocation_queue}
The allocation queue is a ranking of allocation increments assigned to groups. We define $Q_{g,l}\in\left\{1,\dots,{W}\right\}$ as the position of the l'th increment to group $g$ in the queue.
\end{definition}

\begin{theorem}\label{theorem1}
Suppose that Assumption~\ref{assumption:alpha_A} holds and that $\underline{D}_g=0$ for all $g$. Then the solution to the planner's allocation problem, $(D_1^*,...,D_N^*)$, is given by
\begin{align}
D_g^*({W})=\sum_{l=1}^{\overline{D}_g}\mathbbm{1}\left\{Q_{g,l}\leq{W}\right\},
\end{align}
where $Q_{g,l}$ is the position of the l'th increment to group $g$ in the optimal allocation queue:
\begin{equation}\label{eq:optimal_allocation_queue}
Q_{g,l}=\sum_{j=1}^N\sum_{k=1}^{\overline{D}_j}\mathbbm{1}\left\{\frac{\beta_j}{\beta_g}\left(\frac{\left(A_j+\sum_{s=0}^k\alpha_{j,s}\right)^\lambda-\left(A_j+\sum_{s=0}^{k-1}\alpha_{j,s}\right)^\lambda}{\left(A_g+\sum_{s=0}^l\alpha_{g,s}\right)^\lambda-\left(A_g+\sum_{s=0}^{l-1}\alpha_{g,s}\right)^\lambda}\right)\geq1\right\}.
\end{equation}
\end{theorem}
\noindent The proof of Theorem~\ref{theorem1} is in the Appendix.

There are three key parts to Theorem~\ref{theorem1}.\footnote{Theorem~\ref{theorem1} allows for ties and group-aggregation. When the number of candidate recipients of the same type increases, the relative rankings across types are preserved.} First, the ranking of allocation increments along the optimal allocation queue, $Q_{g,l}$, is based on a comparison of the level of group outcomes with and without the next allocation increment. Accordingly, for group $g$'s $l$'th allocation increment, only $\beta_g\left[\left(A_g +\sum_{s=0}^l\alpha_{g,s}\right)^\lambda-\left(A_g +\sum_{s=0}^{l-1}\alpha_{g,s}\right)^\lambda\right]$ needs to be evaluated.\footnote{In the case where $\beta_g=\frac{1}{N}$ and $\lambda=1$, Equation~\eqref{eq:optimal_allocation_queue} simplifies to a descending sort over $\alpha_{g,l}$. Conversely, when $\lambda\rightarrow-\infty$, the allocation queue simplifies to an ascending sort over $A_{g,l}\equiv A_g+\sum_{s=1}^l\alpha_{g,s}$.} Second, while aggregate resources govern what allocations are feasible, the optimal allocation queue is invariant to aggregate resources. That is, the ranking of allocation increments does not vary with the total budget, ${W}$. Third, the optimal allocation queue allows one to compute the resource cost of misallocation, for example measured as the percentage of resources that can be saved by allocating optimally compared to alternative allocations. This measure, which we refer to as \emph{Resource Equivalent Variation} (REV), is given by:
\begin{align}
\label{eq:REV_theorem1}
\begin{gathered}
{W}^o\left(1-REV\right)=\min\left\{{W}:\frac{\sum_{g=1}^N\beta_g\left(A_g+
\sum_{l=0}^{\overline{D}_g}\left(\alpha_{g,l}\mathbbm{1}\left\{Q_{g,l}\leq{W}\right\}\right)\right)^\lambda}{\sum_{g=1}^N\beta_g\left(A_g+\sum_{l=0}^{D_g^o}\alpha_{g,l}\right)^\lambda}\geq1\right\}
\thinspace\thinspace,\\
\end{gathered}
\end{align}
where $\left\{D_g^o\right\}_{g=1}^N$ are alternative allocations, with resources $W^o =\sum_{g=1}^N D_g^o$. REV provides a common scale in resource units by which alternative allocations can be compared with the optimal allocation.\footnote{
In Section~\ref{sec:results_2008_policy}, we use REV to show the potential savings from implementing the optimal policy, keeping the amount of stimulus constant. In Section~\ref{sec:results_2021_policy}, REV could similarly be used to quantify the potential savings from implementing the optimal policy, keeping the government's CES-index over consumer welfare constant.} Similarly, various statistics such as the Gini coefficient of recipient outcomes can be expressed as functions of the optimal allocation queue (see Online Appendix Section~\ref{appendix:sec:Additional_formulas} for details).

Given inputs $\alpha_{g,l}$ and $A_g$ for all $(g,l)$, the computational implementation of Theorem~\ref{theorem1} in the \href{https://fanwangecon.github.io/PrjOptiSNW/}{companion code package} solves for optimal allocation queues along with aggregate outcomes for any value of planner preferences.

\section{Empirical application: stimulus checks}\label{sec:model}
We illustrate our algorithm by means of two applications. In Section~\ref{sec:results_2008_policy}, we derive the optimal allocation of stimulus checks in a setting where the objective of the planner is to stimulate consumption. This application is motivated by the Economic Stimulus Act of 2008, whereby households received stimulus checks in the form of tax rebates in an effort to stimulate the economy in the wake of the Great Recession. In Section~\ref{sec:results_2021_policy}, we derive the optimal allocation of stimulus checks in a setting where the objective of the inequality-averse planner is to maximize welfare. This application is motivated by the CARES Act of 2020 and the ARP Act of 2021, in which the government issued stimulus checks in the form of cash transfers as part of their efforts in insuring individuals against the economic fallout brought about by the COVID-19 pandemic. For the first application, we need to know each household-type $g$'s marginal propensity to consume out of different tax rebate amounts. For the second application, we need to know the households' marginal change in lifetime utility from different check amounts as well as their level of lifetime utility in the case of no allocations. These inputs correspond to household $g$'s  $A_g$ and $\alpha_{g,l}$ values. We obtain these inputs from a life-cycle consumption-savings model with heterogeneous consumers, which predicts household-specific consumption responses to tax rebates and checks. Consumers are forward looking and may save part of these transfers for either retirement, future child expenses, or to buffer future income risk. Single and married consumers have different propensities to save due to different income and family size transition probabilities. The consumption needs of a household  grow less than proportionally with the number of household members due to economies-of-scale.

The following section presents the problem solved by the households during non-crisis periods. Further modeling details about the problem solved by the consumers during the periods of the 2008--2009 and 2020--2021 crises are given in Sections~\ref{sec:results_2008_policy} and~\ref{sec:results_2021_policy}, respectively.

\subsection{Model}\label{Model}

\textit{Households}---The economy is populated by heterogeneous households. The idiosyncratic state of the \emph{household head} (referred to as the \emph{agent}) is denoted by $\omega=\left(j,a,\eta,e,m,k,\delta,\nu\right),$ where $j$ is age, $a$ is non-negative assets, $\eta$ is stochastic labor productivity, $e$ is educational attainment, $m$ is marital status, $k$ is the number of children under age 18, $\delta$ is the discount factor, and $\nu$ is the stochastic labor productivity of the spouse in the event that the agent is married. Both productivity shocks follow Markov processes. Educational attainment is permanent and takes two values: college or non-college. Similarly, marital status is permanent and takes two values: married or single. The number of children under age 18 follows a Markov process that depends on the agent's current number of children, age, educational attainment, and marital status. The discount factor is permanent and takes two values: $\delta\in\{\underline{\delta},\overline{\delta}\}$. Agents retire at age $j_{R}$ and live at most $J$ periods. The probability of survival varies with the agent's age, $\psi_{j}$.

\textit{Income}---Labor productivity varies with the agent's age, educational attainment, and stochastic labor productivity, $\epsilon_{j,\eta,e}$. There is little evidence that individuals reduced their labor-supply in 2008 and during the pandemic in response to the one-time stimulus checks, which is the focus of our analysis. We therefore simplify the analysis by assuming that labor is supplied inelastically.\footnote{As discussed in Sections~\ref{sec:results_2008_policy} and ~\ref{sec:results_2021_policy}, we model the one-time stimulus checks as unexpected policies and there is thus no ``moral hazard'' effects in the model whereby agents internalize how their choices affect eligibility for the stimulus checks. It is beyond the scope of our model to analyze permanent policies of supporting households during economic recessions, for which a more comprehensive consumer model where these moral hazard effects are internalized is required.} Retired agents receive Social Security benefits from the government. To reduce computational costs, we build on \textcite{denardiLifetimeCostsBad2017, nygaardImpactEmployersponsoredInsurance2021} and assume that Social Security benefits are tied to the agent's fixed productivity type as given by their educational attainment, $SS_{e}.$ Spousal income varies with the spouse's labor productivity shock and with the household head's age, educational attainment, number of children, and income, $B_{j,e,k,\eta,\nu}$.

\textit{Government}---The government provides Social Security and consumes goods, $G.$ The latter is included to equalize total government expenditures in the model and the data, thereby ensuring that the tax burden in the model is consistent with the data. The government finances its expenditures by means of progressive income taxes, $T_{y}$, where $y$ is household income.

\textit{Agent's problem}---At time $t$, agents choose how much to consume, $c_t$, and how much to save, $a_{t+1}$. For notational simplicity, we drop time subscripts and use $\prime$ to denote next-period variables. The value function is given by:
\begin{equation}
\begin{array}{lll}
V\left(\omega\right) & = & \underset{c\geq0,a^{\prime}\geq0}{\max}\,\,\,u\left(c,m,k\right)+\delta\psi_{j}\text{\ensuremath{\mathbb{E}}}_{\eta^{\prime}\vert\eta}\text{\ensuremath{\mathbb{E}}}_{\nu^{\prime}\vert\nu}\text{\ensuremath{\mathbb{E}}}_{k^{\prime}\vert\left(j,e,m,k\right)}V\left(\omega^{\prime}\right)\\
 & \text{s.t.} & c+a^{\prime}=a+y-T_{y}\\
 &  & y=ra+\mathbbm{1}_{j<j_{R}}\theta\epsilon_{j,\eta,e}+\mathbbm{1}_{j\geq j_{R}}SS_{e}+\mathbbm{1}_{m=1}B_{j,e,k,\eta,\nu},
\end{array}\label{eq:Value-function}
\end{equation}
where $\theta$ denotes aggregate labor productivity, $r$ is the real interest rate, $\mathbbm{1}_{j<j_{R}}$ $\left(\mathbbm{1}_{j\geq j_{R}}\right)$ are indicator functions that equals one for an agent younger than (at least as old as) $j_{R}$, and $\mathbbm{1}_{m=1}$ is an indicator function that equals one for a married agent.\footnote{Agents in our model are finitely lived, face age-specific mortality risk, and have heterogeneous discount factors. The assumption that agents cannot borrow simplifies our analysis by ruling out the possibility of strategically accumulating debt in anticipation of death. We believe the optimal allocation results presented in Sections~\ref{sec:results_2008_policy} and~\ref{sec:results_2021_policy} are robust to a more comprehensive consumer model whereby agents can borrow but face age-specific borrowing limits.}

\subsection{Calibration}\label{subsec:Calibration}
This section discusses the calibration of the model during non-crisis periods. See Online Appendix Section~\ref{sec:Appendix_Calibration}
for further details. Calibration details for the crisis periods are given in Sections~\ref{sec:results_2008_policy} and~\ref{sec:results_2021_policy}.

\textit{Life-cycle parameters}---Agents enter the economy at age 18, retire at 65, and survive until at most age 100. We use U.S. life-tables for 2020 to obtain age-specific survival probabilities and data from the Panel Study of Income Dynamics (PSID) to derive the probability of being college-educated, the probability of being married, and the initial distribution of children.

Following OECD recommendations, we apply the square-root scale in the model to account for economies-of-scale in consumption. The agent's utility from household consumption $c$ is given by
\begin{equation}
u\left(c,m,k\right)=\frac{\left(\frac{c}{\sqrt{HH\left(m,k\right)}}\right)^{1-\gamma}}{1-\gamma},
\end{equation}
where $HH\left(m,k\right)$ is the number of household members. We set $\gamma$ equal to 2 to match an intertemporal elasticity of substitution of 0.5.

\textit{Technology parameters}---We normalize aggregate productivity, $\theta,$ such that median household income is equal to 1 in the steady state. The interest rate, $r,$ is set to 4 percent per year following \textcite{mcgrattanAverageDebtEquity2003}. Recall that agents have heterogeneous discount factors, $\delta\in\{\underline{\delta},\overline{\delta}\}$. We set $\overline{\delta}$ equal to 0.95 and $\underline{\delta}$ equal to 0.60, where the latter value implies that the consumer strongly prefers to consume rather than save. This is motivated by \textcite{jappelliFiscalPolicyMPC2014}, who show that a version of the Aiyagari model in which a fraction of consumers follow the model-implied optimal consumption rule and the rest follow a rule-of-thumb in which consumption equals income in each period (often labeled ``hand-to-mouth'' behavior) can replicate observed MPC patterns (see also \textcite{aguiarWhoAreHandtomouth2020}). Following \textcite{jappelliIntertemporalChoiceConsumption2006}, we assume that 10 percent of college-educated agents and 40 percent of non-college-educated agents are $\underline{\delta}$-types.

\textit{Transition probabilities for number of children}---We use data from the PSID to derive transition probabilities for the number of children (under age 18) by estimating an ordered logistic regression of the number of children at time $t+1$ conditional on the household head's age, marital status, college attainment, and number of children at time $t$.

\textit{Income}---The labor productivity of an agent of type $\omega$ is given by $\epsilon_{j,\eta,e}=h\left(j,e\right)\exp\left(\eta\right)$, where $h\left(j,e\right)$ is age- and education-specific deterministic labor productivity and $\eta$ is a stochastic labor productivity shock given by
\begin{equation}
\begin{array}{cc}
\eta=\rho\eta_{-1}+\mu, & \mu\sim N\left(0,\sigma_{\mu}^{2}\right).\end{array}
\end{equation}
We use the age- and education-specific life-cycle labor productivity profiles estimated by \textcite{conesaImplicationsIncreasingCollege2020}. Following \textcite{pashchenkoQuantitativeAnalysisHealth2013}, we set the persistence of stochastic productivity shocks, $\rho,$ equal to 0.980 and the variance of the shocks, $\sigma_\mu^{2},$ equal to 0.018.

We use data for married individuals in the PSID to estimate spousal income, $B_{j,e,k,\eta,\nu}$. We regress the logarithm of spousal income on the household head's age, college attainment, number of children, and the logarithm of the household head's income to obtain both the type-specific mean and variance of spousal income.

\textit{Taxes and transfers}---We calibrate Social Security benefits for non-college and college-educated consumers to match the corresponding average benefits in the Current Population Survey. We calibrate government consumption to match data from the Bureau of Economic Analysis (BEA) on the ratio of government consumption expenditures to GDP. Following \textcite{gouveiaEffectiveFederalIndividual1994}, we use the income tax function
\begin{equation}
T_{y}=a_{0}\left(y-\left(y^{-a_{1}}+a_{2}\right)^{-\frac{1}{a_{1}}}\right).
\end{equation}
We use their estimates for $a_{0}$ and $a_{1}$, and adjust $a_{2}$ period-by-period to balance the government budget.

\section{Optimal allocation of stimulus checks: Economic Stimulus Act of 2008}\label{sec:results_2008_policy}
The Economic Stimulus Act of 2008 contained several provisions to boost the economy in the wake of the Great Recession. One of these provisions was a tax rebate which consisted of a basic payment and, conditional on eligibility, a supplemental payment of \$300 per child. To be eligible for the basic payment in 2008, households had to have positive net income tax liability or sufficient qualifying income on their 2007 tax return. As discussed by \textcite{parkerConsumerSpendingEconomic2013a}, the minimum payment for eligible households was  \$300 for single filers (\$600 for married couples filing jointly) or, if larger, the amount of their tax liability with a maximum of \$600 (\$1,200).  The tax rebates phased out with income at a rate of 5 percent for single filers with income exceeding \$75,000 (\$150,000 for married couples filing jointly). This section applies our algorithm to study the optimal allocation of these tax rebates and quantifies the budget savings that the government could have achieved by implementing the optimal policy.

\subsection{Consumer problem during the 2008--2009 Great Recession}\label{sec:Consumer-problem-2008}
We start by describing the problem solved by the consumers and the planner. Data from \textcite{sahmHouseholdResponse20082010} show that nearly all tax rebates were disbursed during the months of May, June, and July of 2008, during which the unemployment rate was still below 6 percent. Motivated by this timing, we model the Great Recession as a two-period shock, where a period in the model is one year. In the first period, agents receive stimulus checks in the form of tax rebates, where both eligibility and the amount received are tied to the household's income and size (and hence tax liability) in the previous (non-crisis) period. In the second period, agents are subject to unemployment risk which varies with their age and educational attainment, $\pi^{U}\left(\omega\right)$. The probability of unemployment, which is assumed to be known to the agents during the first crisis-period, affects the agents' propensity to consume out of the tax rebates. Let $\xi\in[0,1]$ govern the duration of the unemployment spell. Unemployed agents are eligible for UI benefits which replace a share $b\in[0,1]$ of lost earnings.\footnote{We use the average age- and education-specific unemployment probability in 2009 as reported by the Bureau of Labor Statistics (BLS). The duration of the unemployment spell, $1-\xi$, is calibrated to match the average unemployment duration in that year from the BLS. We calibrate the UI replacement rate, $b$, to match the ratio of aggregate UI benefits to aggregate wages and salaries as reported by the BEA in 2009.}

\textit{Agent's problem}---Let $V^{U}\left(\omega\right)$ denote the value function for an agent of type $\omega$ that is unemployed for at least part of the second crisis-period, $\xi<1$, details of which is given in Appendix \ref{sec:Appendix_value_function_details-2008}. Similarly, let $V^{W}\left(\omega\right)$ denote the value function for an agent of type $\omega$ that is employed during the second crisis-period, $\xi=1$. Finally, let $\widehat{V}\left(\omega;D\right)$ denote the value function for an agent of type $\omega$ during the first crisis-period that receives an amount $D\geq0$ in tax rebates, modeled as direct transfers to the agents:
\begin{equation}
\begin{array}{lll}
\widehat{V}\left(\omega;D\right) & = & \underset{c\geq0,a^{\prime}\geq0}{\max}\,\,\,u\left(c,m,k\right)+\delta\psi_{j}\text{\ensuremath{\mathbb{E}}}_{\eta^{\prime}\vert\eta}\text{\ensuremath{\mathbb{E}}}_{\nu^{\prime}\vert\nu}\text{\ensuremath{\mathbb{E}}}_{k^{\prime}\vert\left(j,e,m,k\right)}
\left(
\begin{array}{c}
     \pi^{U}\left(\omega^\prime\right)V^{U}\left(\omega^{\prime}\right) \\
     +(1-\pi^{U}\left(\omega^\prime\right))V^{W}\left(\omega^{\prime}\right)
\end{array}
\right)\\
 & \text{s.t.} & c+a^{\prime}=a+y-T_y+D\\
 &  & y=ra+\mathbb{I}_{j<j_{R}}\theta\epsilon_{j,\eta,e}+\mathbb{I}_{j\geq j_{R}}SS_{e}+\mathbb{I}_{m=1}B_{j,e,k,\eta,\nu}.
\end{array}\label{eq:V_hat-Value-function-2008}
\end{equation}

\subsection{Planner problem during the 2008--2009 Great Recession}\label{sec:Planner-problem-2008}
Let $c\left(\omega;D\right)$ denote consumption of an agent's household of type $\omega$ who receives an amount $D$ in tax rebates. Because the amount received under the actual policy varied with household income and family size (and hence tax liability) as reported on their most recent tax return, we focus on the optimal allocation of 2008 tax rebates given 2007 (pre-crisis) household characteristics. Let $\tilde{c}\left(y,j,m,k;D\right)$ denote the ex-ante expected household consumption in 2008 of an agent with income $y$, age $j$, marital status $m$, and number of children $k$ in 2007 that receives $D$:
\begin{equation}
\label{eq:C_tilde}
\begin{array}{ll}
 & \tilde{c}\left(y,j,m,k;D\right)\\
 = & \psi_{j}\text{\ensuremath{\mathbb{E}}}_{\eta^{\prime}\vert\eta}\text{\ensuremath{\mathbb{E}}}_{\nu^{\prime}\vert\nu}\text{\ensuremath{\mathbb{E}}}_{k^{\prime}\vert\left(j,e,m,k\right)}\ensuremath{\mathbb{E}}_{\delta\vert e}\ensuremath{\mathbb{E}}_{(a,\eta,e,\nu)\vert(y,j,m,k)} 
 c\left(j+1,a'\left(\omega\right),\eta^{\prime},e,m,k^{\prime},\delta,\nu^{\prime};D\right),
\end{array}
\end{equation}
where $a'\left(\omega\right)$ is next period's assets, $\ensuremath{\mathbb{E}}_{\delta\vert e}$ is the expected discount factor given the agent's educational attainment, and $\ensuremath{\mathbb{E}}_{(a,\eta,e,\nu)\vert(y,j,m,k)}$ is the expected value given the probability distribution over assets, stochastic labor productivity, and educational attainment on the agent's income, age, marital status, and number of children.

The planner chooses the amount, $D_g$, for each group $g\in\{1,\dots,G\}$, where groups are defined by marital status, number of children, age, and income.\footnote{The conditioning on age allows us to study optimal age-specific allocations (see our \href{https://fanwangecon.github.io/PrjOptiSNW/}{companion website} for results). Section~\ref{sec:Results-2008} studies non-age-specific optimal allocations.} Let $\Omega_g$ denote the set of agents of type $s=(y,j,m,k)$ that belongs to group $g$:
\begin{equation}
\Omega_g\equiv\left\{s=\left(y,j,m,k\right) : y\in\left[\underline{y}_g,\overline{y}_g\right],j\in\left[\underline{j}_g,\overline{j}_g\right],m=m_g,k=k_g\right\}.
\end{equation}
Let ${W}$ denote the total budget available. The planner's choice set is given by:
\begin{equation}
\mathbb{C}^{D}\equiv\left\{\textbf{D}=\left(D_1,\dots,D_G\right) : D_g\in\left\{\underbar{D}_g,\dots,\overline{D}_g\right\},\sum_{g=1}^{G}\left(\int \mu\left(s\right)\mathbbm{1}_{s\in \Omega_g}ds\cdot D_g\right)\leq{W}\right\},
\label{eq:Choice_set}
\end{equation}
where $\mu(s)$ is the number of agents of type $s$ and $\underbar{D}_g\geq0$ and $\overline{D}_g\geq \underbar{D}_g$ are group-specific lower and upper bounds on transfers to capture that the maximum tax rebates under the actual policy varied with household characteristics such as income and civic status. We then get the following expression for the planner's optimization problem:
\begin{equation}
\underset{\textbf{D}\in\mathbb{C}^{D}}{\max} \int \tilde{c}\left(s;\sum_{g=1}^{G} D_g\cdot\mathbbm{1}_{s\in\Omega_g}\right)\mu\left(s\right)ds.
\label{eq:Planner_objective_emp_app-2008}
\end{equation}

\subsection{Optimal consumption stimulus allocation}\label{sec:Results-2008}
The consumption stimulus effect will be maximized for a given budget by allocating the tax rebates to consumer-groups with the highest marginal propensity to consume (MPC)---taking into account that as rebates get larger, the MPC declines.\footnote{The consumption stimulus planner corresponds to a planner with $\lambda=1$.} This will typically imply that rebates are given to consumers with a priori low consumption, as we illustrate in Figure~\ref{fig:aveC_vs_MPC}. This figure shows on the X-axis  average consumption per household member (for brevity, referred to as per-capita consumption below) before receiving tax rebates and  on the Y-axis the corresponding MPCs out of the first \$100 in tax rebates for different demographic groups: single or married with 0--4 children (averaged over other idiosyncratic states such as age and educational attainment using population weights). The size of each marker corresponds to the group's relative population weight.

We see that for all groups the MPCs decline steeply with per-capita consumption, but that the slope varies across groups. Whereas MPCs decline near-linearly with average per-capita consumption for married households, the slope is non-linear for singles without children who account for the largest fraction of the population, in particular at low to moderate consumption levels. The MPCs for a given level of per-capita consumption are lower for married than single households, which is partially due to the two group's different family size transition probabilities. In particular, married households are more likely to have a child in the near future, following which they will have more mouths to feed and lower average spousal income (see Online Appendix Section~\ref{sec:Appendix_Calibration} for details).

The queue for the optimal consumption stimulus allocation is based on a ranking of MPCs. In Table~\ref{table:APCandMPC}, we illustrate the queue as it would look for coarser income categories and fewer household types (see the \href{https://fanwangecon.github.io/PrjOptiSNW/}{companion website} for the full MPC table). For each household type, the planner needs to know the entire ``path'' of MPCs as the size of the tax rebate increases. The top panel of Table~\ref{table:APCandMPC} displays, in the left-most column, predicted initial per-capita consumption before receiving tax rebates and, in the right-most columns, MPCs out of an \emph{additional} \$100 increase in tax rebates for different household types: single or married with 0 or 2 children and different household income levels (averaged over other idiosyncratic states using population weights).\footnote{The columns report the expected value of $MPC^{\$100}(g; D) = \frac{\Delta c(g)}{(D+\$100)-D}$, where $\Delta c(g)$ is the change in consumption of a household of type $g$ whose transfer increases from $D\geq0$ to $D+\$100$.} For example, the column labeled \$1,000 shows the MPC for a household that has already received \$1,000 in tax rebates, measured as the amount by which consumption will increase if the household's tax rebate increases from \$1,000 to \$1,100.

Table~\ref{table:APCandMPC} shows that the MPC for singles without children making less than \$20,000 per year receiving a tax rebate of \$100 takes a value of 60.2 percent. In the same row, the column labeled \$3,000 shows an MPC of 48.1 percent, which implies that the consumption of this household type is predicted to increase by an additional \$48.1 if the tax rebate increases from \$3,000 to \$3,100 (``the MPC when receiving \$3,000''). MPCs for higher-income households are much smaller for small tax rebates, but decline more slowly with the amount, and are also higher for households with more children. 
It is also evident that for given household income and number of children, the MPCs for the smallest checks for married households are lower than for corresponding singles. This follows because couples are older on average and because they are more likely to have children in the near future. However, for married households the MPCs decline relatively slower with the size of the tax rebate.\footnote{The decline in MPCs for, say, singles with 2 children, is initially very steep followed by a flat segment when tax rebates increase from \$500 to \$3,000. This pattern is largely due to the coarseness of the asset grid and MPCs decline with income in the underlying finer categories from which the MPCs in Table~\ref{table:APCandMPC} are generated as averages. For the optimal allocations, we consider no less than 67,070,640 MPCs, see Technical Appendix Section~\ref{sec:Appendix_computational_details}.}

The MPC panel illustrates our search for the allocation that maximizes the stimulus effect on consumption. For example, a single with two children and income in the \$40,000--\$60,000 range tentatively assigned a \$3,000 tax rebate will be queued ahead of a single without children and income in the \$0--\$20,000 range tentatively assigned a tax rebate of \$2,000 because the \$100 increment results in a consumption increase of \$52.2 for the former and \$48.3 for the latter.

The bottom panel of Table~\ref{table:APCandMPC} shows the average propensity to consume (APC) for the same household types conditional on the transfer amount.\footnote{The columns report the value of $APC(g; D) = \frac{\Delta c(g)}{D}$, where $\Delta c(g)$ is the change in consumption of a household of type $g$ who receives a tax rebate of $D>0$.}  As these are averages of the MPCs, they decline more slowly with transfer amounts. For example, for singles without children in the lowest income category, they decline from 60.2 to 53.1 percent as the transfer amount increases from \$100 to \$3,100, compared to a corresponding MPC reduction from 60.2 to 48.1 percent. What matters for the optimal stimulus allocation is each household type's path of MPCs, not the APCs that are more commonly observed in data.\footnote{One exception is \textcite{fagerengMPCHeterogeneityHousehold2021}, who estimate MPCs out of different lottery amounts using Norwegian register data.}

Figure~\ref{fig:optimal2008} compares the actual allocation to the optimal consumption stimulus allocation, assuming the government increased the maximum tax rebate amount to \$900 per adult and the maximum supplemental child payment, conditional on eligibility, to \$600 per child.\footnote{
More formally, the tax rebate limit for single filers is now the maximum of \$300 and their tax liability, with a maximum of \$900 if their tax liability exceeds \$300, plus a supplemental payment of up to \$600 per child. The tax rebate limit for married couples filing jointly is similarly set at the maximum of \$600 and their tax liability, with a maximum of \$1800 if their tax liability exceeds \$600, plus a supplemental payment of up to \$600 per child.
} Not surprisingly, the optimal allocation is tilted more towards the poorest (due to their higher MPCs) who receives the maximum amount (to the extent that their tax liability is as large). However, most of the reallocation from the actual rebate takes the form of larger rebates to singles with children.
Because we evaluate the optimal allocation for a given budget, the money needs to be reallocated from married couples due to their relatively lower MPCs, for which only the lowest income groups receive the maximal amount, while higher-income households would receive no tax rebates. The phase-out occurs around \$50,000 for couples with 0--1 children and around \$70,000 for couples with 2 children. Whether this optimal pattern would have been politically feasible or not is beyond the scope of our paper to evaluate.

In Table~\ref{table:REV}, we report the REV corresponding to different allocation constraints, measured as the percentage by which the government can reduce total spending on tax rebates and still achieve the same aggregate consumption stimulus. For the actual 2008 maximum tax rebate limits, the government could have obtained the same stimulus with a 2.0 percent lower budget if it allocated optimally. However, significant savings of more than 10.0 percent could have been obtained by allocating optimally if the tax rebate limits per adult and/or per child were increased.

\section{Optimal allocation of stimulus checks: CARES Act of 2020 and American Rescue Plan Act of 2021}\label{sec:results_2021_policy}
Following the worldwide outbreak of COVID-19 infections, countries around the world responded by closing down businesses for extended periods of time and pumping out unprecedented amounts of money to ameliorate the adverse economic effects of the pandemic. In the U.S., the American Rescue Plan (ARP) Act of March 2021 had a budget of 1.9 trillion dollars, about \$400 billion of which was spent on direct checks to households as part of its efforts in insuring individuals against the economic fallout.\footnote{Direct checks to households were also included in the CARES Act of March 2020 (maximum of \$1,200 per adult) and the Consolidated Appropriations Act of December 2020 (maximum of \$600 per adult).} The actual allocation was a \$1,400 check to non-married individuals with income less than \$75,000 and a \$2,800 check to married couples if their joint income was less than \$150,000 (gradually phased out at higher income levels), with an additional \$1,400 per dependent. This section applies our algorithm to study the optimal allocation of these checks in a setting where an inequality-averse planner maximizes the welfare of check recipients.

\subsection{Consumer problem during the 2020--2021 pandemic}\label{sec:Consumer-problem-2021}
As for the previous application, we start by describing the problem solved by the consumers and the planner. We model COVID-19 as a two-period shock. In the first period, agents are subject to an unexpected unemployment shock. Consistent with actual policies (CARES Act and Consolidated Appropriations Act), eligible agents receive up to two rounds of checks and the generosity of unemployment insurance (UI) benefits increases.\footnote{The increased generosity of UI benefits was designed to replace 100 percent of average U.S. wages. \textcite{ganongUSUnemploymentInsurance2020} show that the median replacement rate was 134 percent and about two-thirds of eligible unemployed workers received benefits exceeding their lost earnings.} In this section, we focus on the optimal allocation of the \emph{third} round of checks, which are allocated during the second crisis-period in the model (the third round of checks refers to the checks allocated under the ARP Act). Given that the implications of the first crisis-period are endogenously captured by the agents' assets at the start of the second period, we only present the problem solved by the agents during the second period of COVID-19.

Using administrative payroll data, \textcite{cajnerLaborMarketBeginning2020} document that younger, older, and low-income workers were more likely to lose their job during the pandemic. Given this, we model the surge in unemployment as an unexpected unemployment shock and assume that the probability of unemployment varies with the agent's age and earnings. As in Section~\ref{sec:Consumer-problem-2008}, let $\pi^{U}\left(\omega\right)$ denote the unemployment probability of an agent of type $\omega$, let $\xi\in[0,1]$ govern the duration of the unemployment spell, and let $b\in[0,1]$ denote the UI replacement rate.\footnote{We adjust the 2020 unemployment probabilities to match the April 2020 estimates in \textcite{cajnerLaborMarketBeginning2020}. We let the corresponding 2021 probabilities be given by their June 2020 estimates, but scaled down by a common factor to match the aggregate unemployment rate in February 2021. Given the increased UI replacement rate during the pandemic, we assume that UI replaces 100 percent of lost earnings, which means that the duration of the unemployment spell, $\xi$, does not affect income in the benchmark analysis. Sensitivity analyses reported in Appendix Section~\ref{sec:Appendix_additional_results} show that our results are not sensitive to the choice of the UI replacement rate, $b$.}

The large reduction in consumption brought about by the pandemic was partially driven by the fact that individuals reduced their consumption due to lockdown of establishments, restrictions on travel, and fear of contracting the virus. \textcite{carrollModelingConsumptionResponse2021} estimate that 10.9 percent of the goods become highly undesirable or unavailable during the pandemic. We account for this by reducing the households' marginal utility of consumption during the pandemic by a scaling factor, $\kappa$, calibrated to match the 10.9 percent reduction in aggregate consumption solely due to the lockdown.

\textit{Agent's problem}---Let $V^{U}\left(\omega;D\right)$ denote the value function for an agent that is unemployed for at least part of the second COVID-19 period, $\xi<1$, and that receives an amount $D\geq0$ in checks, modeled as direct cash transfers to the agents:\footnote{Because the shock is transitory, the economy will transition back to the pre-COVID equilibrium, where the value function is as given in Equation~\eqref{eq:Value-function}.}
\begin{equation}\begin{array}{lll}
V^{U}\left(\omega;D\right) & = & \underset{c\geq0,a^{\prime}\geq0}{\max}\,\,\,\kappa u\left(c,m,k\right)+\delta\psi_{j}\text{\ensuremath{\mathbb{E}}}_{\eta^{\prime}\vert\eta}\text{\ensuremath{\mathbb{E}}}_{\nu^{\prime}\vert\nu}\text{\ensuremath{\mathbb{E}}}_{k^{\prime}\vert\left(j,e,m,k\right)}V\left(\omega^{\prime}\right)\\
 & \text{s.t.} & c+a^{\prime}=a+y-T_y+D\\
 &  & y=ra+\mathbb{I}_{j<j_{R}}\theta\epsilon_{j,\eta,e}\left[\xi+b\left(1-\xi\right)\right]+\mathbb{I}_{j\geq j_{R}}SS_{e}+\mathbb{I}_{m=1}B_{j,e,k,\eta,\nu}.
\end{array}\label{eq:V_U-Value-function-2021}
\end{equation}
Consistent with the CARES Act, Consolidated Appropriations Act, and ARP Act, the checks are exempt from income taxation.\footnote{The implications of the stimulus programs depend on the timing of the financing of these programs. The benchmark analysis assumes that the government never increases taxes to finance their increased expenditures on UI benefits and cash checks; stimulus programs are tantamount to ``manna-from-heaven.'' The optimal allocation results in Section~\ref{sec:Results-2021} are robust to an alternative environment where tax rates are adjusted to balance the government budget period-by-period during the crisis.} Analogously, let $V^{W}\left(\omega;D \right)$ denote the value function for an agent of type $\omega$ that receives an amount $D$ in checks and that is employed during the second COVID-19 period. These agents solve the same problem as described in Equation~\eqref{eq:V_U-Value-function-2021}, with the exception that  $\xi=1$.

\subsection{Planner problem during the 2020--2021 pandemic}\label{sec:Planner-problem-2021}
We want to derive the optimal allocation when the planner's objective is to maximize a CES index over the welfare of check recipients. The value functions for consumers' expected discounted lifetime utility are not necessarily positive, which precludes using value functions directly when  $\lambda\in(-\infty,1)$. We therefore translate consumer welfare into a positive consumption equivalent  following an approach similar to \textcite{benabouTaxEducationPolicy2002, bakisTransitionalDynamicsOptimal2015, boarEfficientRedistribution2020a}. In particular, we first convert the household's lifetime utility, $V^{q}\left(\omega;D\right)$, where $q\in\{U,W\}$, into more interpretable units by deriving the positive constant consumption stream, $x^{q}\left(\omega;D\right)$, that the household would need to receive every period to achieve the same level of lifetime utility:
\begin{equation}\label{eq:constant-cons-stream-V_q}
V^{q}\left(\omega;D\right)=\sum_{\tilde{j}=j(\omega)}^{J}\delta^{\tilde{j}-j(\omega)}\Psi_{\tilde{j}}\frac{x^{q}\left(\omega;D\right)^{1-\gamma}}{1-\gamma},\,\,\,q\in\{U,W\},
\end{equation}
where $\Psi_{\tilde{j}}$ is the likelihood of surviving from age $\tilde{j}=j(\omega)$ to age $J$.

As $x^{q}\left(\omega;D\right)$ is a constant, we can simplify the notation and write $x^{q}\left(\omega;D\right)=\left[\frac{V^{q}\left(\omega;D\right)}{\Lambda(\omega)}(1-\gamma)\right]^{\frac{1}{1-\gamma}}$ for an appropriately defined $\Lambda(\omega)$. But because agents  have heterogeneous discount factors and different expected lifetimes, $x^{q}\left(\omega;D\right)$ generally fails to preserve the ordering of $V^{q}(\omega,D)$ across types $\omega$.\footnote{That is, if $V^{q}(\omega_1;D)>V^{q}(\omega_2;D)$, it is not necessarily the case that $x^{q}(\omega_1;D)>x^{q}(\omega_2;D)$.} To correct for this, we calculate a life-span-adjusted consumption equivalent (which adjusts for expected lifetime and discount factors) as follows:
\begin{equation}\label{eq:certainty-equivalent-welfare-measure}
\overline{x}^{q}(\omega;D)=x^{q}(\omega;D)\Lambda(\omega)^{\frac{1}{1-\gamma}}=\left[V^{q}(\omega;D)(1-\gamma)\right]^{\frac{1}{1-\gamma}},\,\,\,\,q\in\{U,W\}.
\end{equation}
The life-span-adjusted consumption equivalent preserves the ordinal rank of $V^{q}(\omega;D)$ across types $\omega$ for all $\gamma$ (see Appendix Section~\ref{sec:Appendix_computational_details} for the proof).

The amount received under the ARP Act varied with household income and family size as reported on their most recent tax return. We therefore focus on the optimal allocation of 2021 checks given 2020 household characteristics. Let $\tilde{x}\left(y,j,m,k;D\right)$ denote the ex ante expected welfare in 2021 of agents with income $y$, age $j$, marital status $m$, and number of children $k$ in 2020 that receive $D$, defined analogously to Equation~\eqref{eq:C_tilde} (see Appendix Section~\ref{sec:Appendix_value_function_details-2021} for details). The planner's optimization problem is then given by
\begin{equation}
\underset{\textbf{D}\in\mathbb{C}^{D}}{\max} \left(\int\beta_s \tilde{x}\left(s;\sum_{g=1}^{G} D_g\cdot\mathbbm{1}_{s\in\Omega_g}\right)^{\lambda}\mu\left(s\right)ds\right)^{\frac{1}{\lambda}},
\label{eq:Planner_objective_emp_app-2021}
\end{equation}
where the choice set is as given in Equation~\eqref{eq:Choice_set}. The planner weights, $\beta_s$, can be adjusted for whether the planner focuses on households or per-capita when deciding on allocations, and in this application, we choose $\beta_s$ equal to the number of household members.

\subsection{Optimal welfare allocation}\label{sec:Results-2021}
\begin{figure}
\centering
\includegraphics[width=\linewidth]{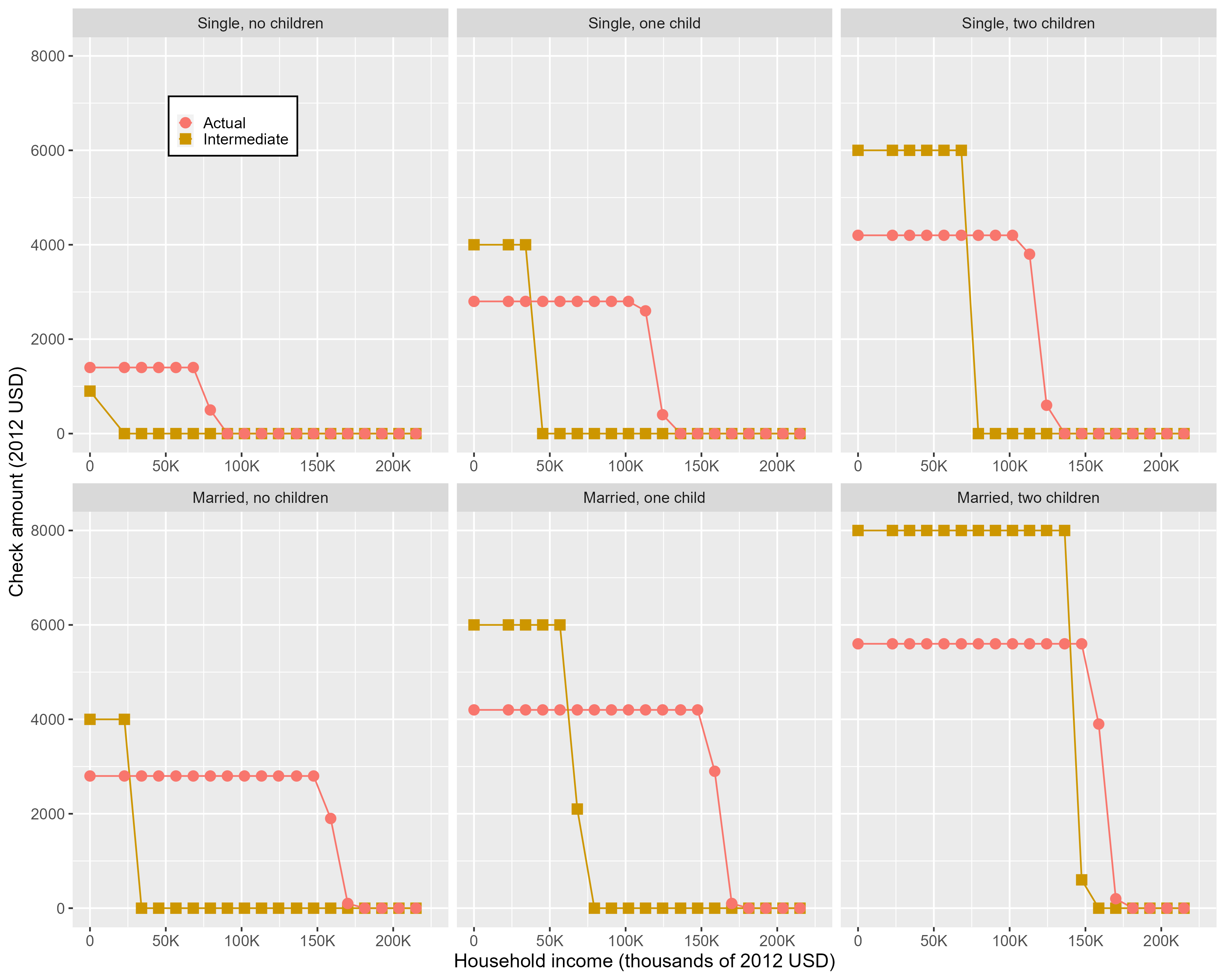}
\smallskip
\caption{Actual vs. Optimal Welfare Allocation by Income and Family Status for Planner with Intermediate Inequality Aversion. Maximum Check Size $\$2$,000 per adult and $\$2$,000 per child.\\
\emph{Notes:} The figure shows the allocation of checks by household income and family status (marital status and number of children) for the 2021 checks and for the optimal welfare allocation of a planner with intermediate inequality aversion ($\lambda=-1$) of the same amount of money calculated under the assumption that the maximum check amount is $\$2,000$ per adult and $\$2,000$ per child.}
\label{fig:covid-intermediate}
\end{figure}

Figure~\ref{fig:covid-intermediate} displays the actual versus optimal checks when the planner cares about the welfare of check recipients, assuming the government has intermediate aversion to welfare inequality ($\lambda=-1$). While the actual allocations were capped at \$1,400 per individual (including dependents), Figure~\ref{fig:covid-intermediate} displays the optimal allocation assuming a cap of \$2,000 per individual.

Compared to the consumption stimulus case studied in Section~\ref{sec:Results-2008}, the most striking difference in the 2021 optimal welfare allocation is the tilt towards families (singles or couples) with several children. In particular, married couples with two children and income less than \$150,000 receive the maximum amount of \$8,000. Because we assume the same budget as the actual allocation, the reallocation towards larger families requires stricter income limits for families with one child or less; for example, a married family with no children would receive no check if their household income was above \$30,000, and among singles without children only the very poorest would receive any check.

\begin{figure}
\centering
\includegraphics[width=\linewidth]{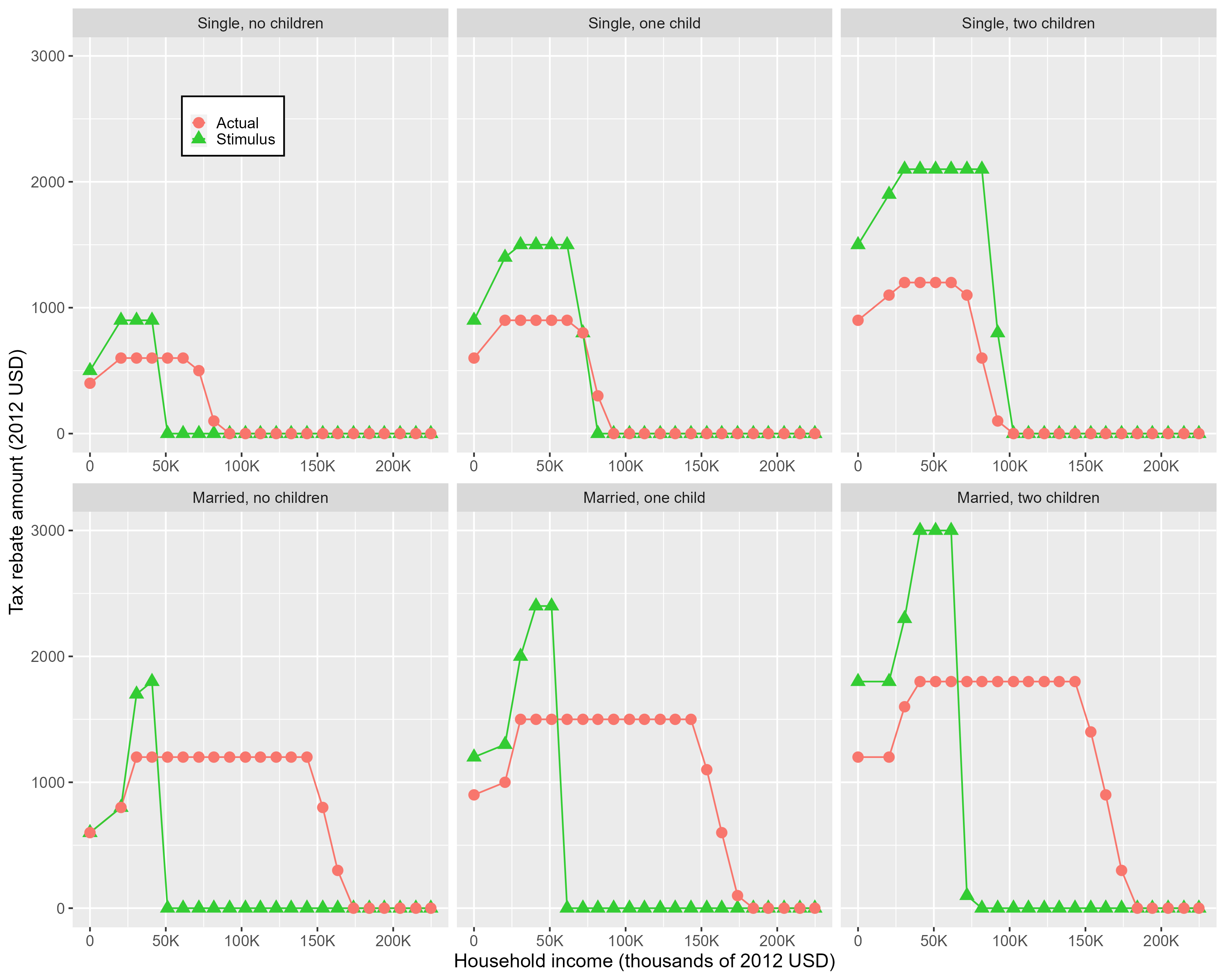}
\smallskip
\caption{Actual vs. Optimal Consumption Stimulus Allocation by Income and Family Status. Maximum Tax Rebate $\$$900 per adult and $\$$600 per child.\\
\emph{Notes:} The figure shows the allocation of tax rebates by household income and family status (marital status and number of children) for the 2008 policy and for the optimal consumption stimulus allocation ($\lambda=1$) of the same amount of money calculated under the assumption that the maximum tax rebate is $\$900$ per adult and $\$600$ per child.}
\label{fig:optimal2008}
\end{figure}

During recessions, the government may worry about both stimulus and welfare of the poorest. However, the optimal allocation of direct transfers to households can be quite different depending on the objective of the government, as can be seen by comparing Figure~\ref{fig:optimal2008} and Figure~\ref{fig:covid-intermediate}. Note, however, that the figures are not directly comparable due to the particular features of the 2008--2009 and 2020--2021 crises, although a qualitatively similar difference would show if the optimal stimulus allocations were displayed for 2021.

\subsection{Deriving confidence bands}\label{sec:Results-perturbation}
If the outcomes for each group are measured with errors, for instance due to estimation uncertainty, the allocations will inherit this uncertainty because it affects the underlying $A_g$ and $\alpha_{g,l}$ used to derive the optimal allocation queue (see Section~\ref{subsec:Solution_to_discrete_allocation_problem}). An advantage of our algorithm is that it can be used to derive confidence bands for the optimal allocations as long as the degree of uncertainty about the inputs can be quantified and the uncertainty preserves the property of non-increasing returns to increments, which will typically hold if the uncertainty is a result of minor modeling choices or due to parameter estimation. This follows because the computational burden required to solve for the optimal allocation queue increases only linearly with the number of recipients, thereby making it computationally feasible to solve the allocation problem for a large number of permutations of the inputs.

We illustrate this by considering the case where each household group's welfare equals the calculated value from the benchmark analysis plus a mean-zero normally distributed error (independent across groups) with standard deviation equal to 10 percent of the base level. We draw a vector of errors (one error for each group) and derive the optimal allocation. We repeat this 500 times and record the allocations to each group.

\begin{figure}
\centering
\includegraphics[width=\linewidth]{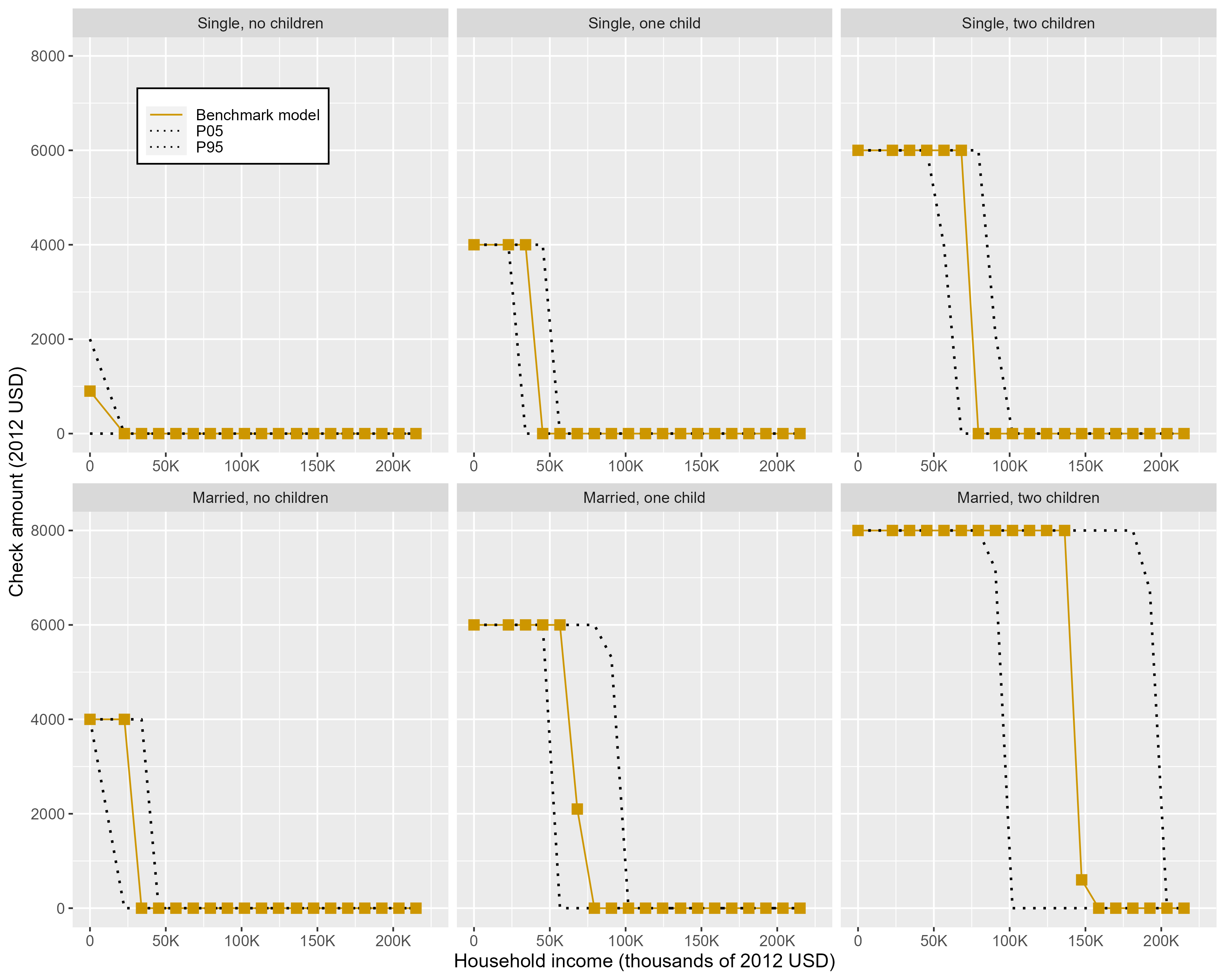}
\smallskip
\caption{Uncertainty of Optimal Allocations when 2021 Welfare is Measured with Error.\\
\emph{Notes:} The solid lines show the optimal welfare allocations (by income and family status for the case with intermediate planner inequality version ($\lambda=-1$) and a maximum check limit of $\$2,000$ per adult and child) under the benchmark parameterization. The dotted lines show the 95 percent ``confidence band'' when welfare for each household type is subject to mean-zero normally distributed errors with a standard deviation equal to 10 percent of group-specific welfare pre-checks.}
\label{fig:pertubations}
\end{figure}

In Figure~\ref{fig:pertubations}, we re-display the optimal allocations from Figure~\ref{fig:covid-intermediate} together with a 95 percent confidence band across the 500 different sets of optimal allocations. The overall pattern of allocating more to poorer households remains, as does the pattern of allocating more to families with several children. However, there is substantial uncertainty in the particular income-limit whereby the checks start to phase out.

While the uncertainty underlying this particular example is not rigorously derived from estimation, it illustrates how our algorithm easily can be used to derive confidence bands for the optimal allocations if the uncertainty in the underlying inputs from the first step of our approach can be quantified.

\section{Conclusion}\label{sec:conclusion}
This paper considers optimal allocation of discrete resources to heterogeneous groups, $g\in\{1,\dots,N\}$, assuming the planner has a fixed budget and CES preferences over the resulting outcomes, $H_{g}(D_g)$. A closed-form solution/algorithm, characterized by an optimal allocation queue, is provided under the assumption that group-specific outcomes $H_{g}$ increase with group-specific allocations at non-increasing rates. The algorithm is applied to allocations of ``support checks'' from the U.S. government to households during the Great Recession and during the COVID-19 pandemic. To obtain the inputs for the analysis, we develop a life-cycle consumption-savings model with heterogeneous households, which predicts household-type-specific consumption and welfare responses to tax rebates and cash transfers. The algorithm can be used to examine the sensitivity of the allocations to different planner objectives and to different degrees of planner inequality-aversion, and we show how the algorithm can be used to derive confidence bands for the optimal allocations when there is uncertainty about the underlying inputs due to for example parameter uncertainty.

The computational demands required to solve for the optimal allocation queue grow linearly with the number of groups, despite the number of possible allocations growing combinatorially, and one can apply our algorithm in cases where, for example, Lagrangian methods become infeasible. We do not focus on inference, but the analytical expression for the optimal allocation queue allows researchers to efficiently compare optimal allocations under different planner preferences, making it feasible to also estimate planner preference parameters that can rationalize observed allocations.

\clearpage
\pagebreak

\section*{Appendix}
\label{Appendix:Proof}
\subsection*{Proof of Theorem 1}

The following proof shows that under Atkinson-CES preference aggregation and non-increasing marginal benefits of incremental allocations, a sequentially local solution is both globally optimal and invariant to aggregate resources. For notational simplicity, we ignore the possibility of ties between candidate recipients. However, Theorem~\ref{theorem1} allows for ties, and the proof presented below generalizes to those cases.
\begin{proof}
The proof of Theorem~\ref{theorem1} is iterative.
Suppose the first $\widetilde{W}$ increments have been allocated. Let $\textbf{D}^{\widetilde{W}}$ denote the corresponding allocations, with $D_i^{\widetilde{W}}$ units allocated to individual $i$.
Consider the problem of maximizing the planner's objective function on the following constraint set
\begin{equation}\label{theorem1proof_choiceset}
\mathbb{C}(\textbf{D}^{\widetilde{W}})\equiv\left\{(D_1,\dots,D_N):D_i\in\{0,1\},D_i+D_i^{\widetilde{W}}\leq\overline{D}_i,\sum_{i=1}^ND_i\leq1)\right\},
\end{equation}
where the planner allocates one unit of resource given existing allocations  $\textbf{D}^{\widetilde{W}}.$ To maximize the planner's objective, the planner allocates the next increment to agent $i$ (which increases the total allocation to $\widetilde{W}+1$) iff, for all $j$ where $D_j^{\widetilde{W}}<\overline{D}_j$, the following holds:
\begin{equation}\label{theorem1proof_part1}
\left(\beta_i\left(A_i+\sum_{l=0}^{D_i^{\widetilde{W}}+1}\alpha_{i,l}\right)^\lambda+\sum_{k=1}^N\beta_k\left(A_k+\sum_{l=0}^{D_k^{\widetilde{W}}}\alpha_{k,l}\right)^\lambda-\beta_i\left(A_i+\sum_{l=0}^{D_i^{\widetilde{W}}}\alpha_{i,l}\right)^\lambda\right)^{\frac{1}{\lambda}}\geq
\end{equation}
\begin{equation*}
\left(\beta_j\left(A_j+\sum_{l=0}^{D_j^{\widetilde{W}}+1}\alpha_{j,l}\right)^\lambda+\sum_{k=1}^N\beta_k\left(A_k+\sum_{l=0}^{D_k^{\widetilde{W}}}\alpha_{k,l}\right)^\lambda-\beta_j\left(A_j+\sum_{l=0}^{D_j^{\widetilde{W}}}\alpha_{j,l}\right)^\lambda\right)^{\frac{1}{\lambda}}.
\end{equation*}
Given $\lambda\in(-\infty,1]$, Equation \eqref{theorem1proof_part1} can be rewritten as
\begin{equation}
\label{theorem1proof_part2}
\resizebox{1.00\hsize}{!}{
\begin{tabular}{@{}l@{}}
$\beta_i\left(A_i+\sum_{l=0}^{D_i^{\widetilde{W}}+1}\alpha_{i,l}\right)^\lambda-\beta_i\left(A_i+\sum_{l=0}^{D_i^{\widetilde{W}}}\alpha_{i,l}\right)^\lambda\geq\beta_j\left(A_j+\sum_{l=0}^{D_j^{\widetilde{W}}+1}\alpha_{j,l}\right)^\lambda-\beta_j\left(A_j+\sum_{l=0}^{D_j^{\widetilde{W}}}\alpha_{j,l}\right)^\lambda\thinspace\text{if}\thinspace\thinspace\thinspace0<\lambda\leq1$\\
$\beta_i\left(A_i+\sum_{l=0}^{D_i^{\widetilde{W}}+1}\alpha_{i,l}\right)^\lambda-\beta_i\left(A_i+\sum_{l=0}^{D_i^{\widetilde{W}}}\alpha_{i,l}\right)^\lambda\leq\beta_j\left(A_j+\sum_{l=0}^{D_j^{\widetilde{W}}+1}\alpha_{j,l}\right)^\lambda-\beta_j\left(A_j+\sum_{l=0}^{D_j^{\widetilde{W}}}\alpha_{j,l}\right)^\lambda\thinspace\text{if}\thinspace\thinspace\thinspace\lambda\leq0.$
\end{tabular}
}
\end{equation}
Under Assumption~\ref{assumption:alpha_A}, $A_i+\alpha_{i,l}>\alpha_{i,l}>0$. Equation~\eqref{theorem1proof_part2} can therefore be rewritten as
\begin{equation}\label{theorem1proof_part3}
D_i^*\left(\textbf{D}^{\widetilde{W}}\right)=1\text{ iff }\frac{\beta_i}{\beta_j}\left(\frac{\left(A_i+\sum_{l=0}^{D_i^{\widetilde{W}}+1}\alpha_{i,l}\right)^\lambda-\left(A_i+\sum_{l=0}^{D_i^{\widetilde{W}}}\alpha_{i,l}\right)^\lambda}{\left(A_j+\sum_{l=0}^{D_j^{\widetilde{W}}+1}\alpha_{j,l}\right)^\lambda-\left(A_j+\sum_{l=0}^{D_j^{\widetilde{W}}}\alpha_{j,l}\right)^\lambda}\right)\geq1.
\end{equation}

Note that the optimality of the first $\widetilde{W}$ allocation increments is unaffected by allocation increment number $\widetilde{W}+1$ due to non-increasing marginal benefits, $\alpha_{i,l}\leq\alpha_{i,l-1}$. The planner's problem therefore simplifies to a local comparison of each individual's marginal benefit of receiving allocation increment number $\widetilde{W}+1$ conditional on the increments that have already been allocated, $\textbf{D}^{\widetilde{W}}.$ Given this, Equation \eqref{theorem1proof_part3} can be rewritten as:
\begin{equation}\label{theorem1proof_part4}
D_i^*\left(\textbf{D}^{\widetilde{W}}\right)=1\text{ iff }\sum_{j=1}^N\sum_{k=0}^{\overline{D}_j}\mathbbm{1}\left\{\frac{\beta_j}{\beta_i}\left(\frac{\left(A_j+\sum_{l=0}^k\alpha_{j,l}\right)^\lambda-\left(A_j+\sum_{l=0}^{k-1}\alpha_{j,l}\right)^\lambda}{\left(A_i+\sum_{l=0}^{D_i^{\widetilde{W}}+1}\alpha_{i,l}\right)^\lambda-\left(A_i+\sum_{l=0}^{D_i^{\widetilde{W}}}\alpha_{i,l}\right)^\lambda}\right)\geq1\right\}=\widetilde{W}+1.
\end{equation}
Each increment of the optimal allocation queue is then given by:
\begin{equation}\label{theorem1proof_part5}
Q_{i,l}=\min\left\{\widetilde{W}:\widetilde{W}\in\left\{1,\dots,\sum_{i=1}^N\overline{D}_i\right\}\text{ and }D_i^{\widetilde{W}}=l\right\},
\end{equation}
which is equal to the $\widetilde{W}$ units of resources required for individual $i$ to optimally receive his or  her $l$'th allocation increment. Equation~\eqref{theorem1proof_part5} is equivalent to the summation in Equation~\eqref{eq:optimal_allocation_queue}.

The iterative optimal solution $Q_{i,l}$ is only optimal if the resource expansion path does not bend backward.
This means that if it is optimal to allocate consecutively to $i$, it is also optimal to allocate to $i$ over $\mathbb{C}(\textbf{D}^{\widetilde{W}})$ iteratively. Without loss of generality, suppose $\widetilde{W}=0$, the requirement is then, $\forall j$,
\begin{align}
  \begin{split}
    \label{eq:pfdise}
    \text{if}
          \left(
            \frac{
                  \left(A_{i} + \alpha_{i,1} + \alpha_{i,2} \right)^{\lambda}
                  -
                  \left(A_{i} + \alpha_{i,1} \right)^{\lambda}
                  }{
                  \left(A_{j} + \alpha_{j,1} \right)^{\lambda}
                  -
                  \left(A_{j}  \right)^{\lambda}
                  }
          \right)
              \ge
          \frac{\beta_{j}}{\beta_{i}}
    \text{, then}
          \left(
            \frac{
                  \left(A_{i} + \alpha_{i,1} \right)^{\lambda}
                  -
                  \left(A_{i} \right)^{\lambda}
                  }{
                  \left(A_{j} + \alpha_{j,1} \right)^{\lambda}
                  -
                  \left(A_{j}  \right)^{\lambda}
                  }
          \right)
              \ge
          \frac{\beta_{j}}{\beta_{i}}.
  \end{split}
\end{align}

Assumption $\alpha_{i,l-1} \ge \alpha_{i,l}$ satisfies Equation~\eqref{eq:pfdise}. For the inequalities in Equation~\eqref{eq:pfdise}, comparisons are based on differences in the numerators. Let $g(A ; \alpha) = \left( A + \alpha \right)^{\lambda} - \left( A \right)^{\lambda}$. Given $\alpha_{i,1} \ge \alpha_{i,2}$, if $0<\lambda<1$, then $0 < g(A_{i} + \alpha_{i,1} ; \alpha_{i,2}) \le g(A_{i} + \alpha_{i,1} ; \alpha_{i,1}) < g(A_{i} ; \alpha_{i,1})$. This satisfies the condition in Equation~\eqref{eq:pfdise}. If $\lambda<0$, both the numerator and the denominator are negative. We have, given $\alpha_{i,1} \ge \alpha_{i,2}$, $0 > g(A_{i} + \alpha_{i,1} ; \alpha_{i,2}) \ge g(A_{i} + \alpha_{i,1} ; \alpha_{i,1}) > g(A_{i} ; \alpha_{i,1})$. This also satisfies the condition in Equation~\eqref{eq:pfdise}.
\end{proof}

\clearpage

\begingroup
\setstretch{1.0}
\setlength\bibitemsep{3pt}
\printbibliography[title=References]
\endgroup
\pagebreak


\begin{figure}
\centering
\includegraphics[width=\linewidth]{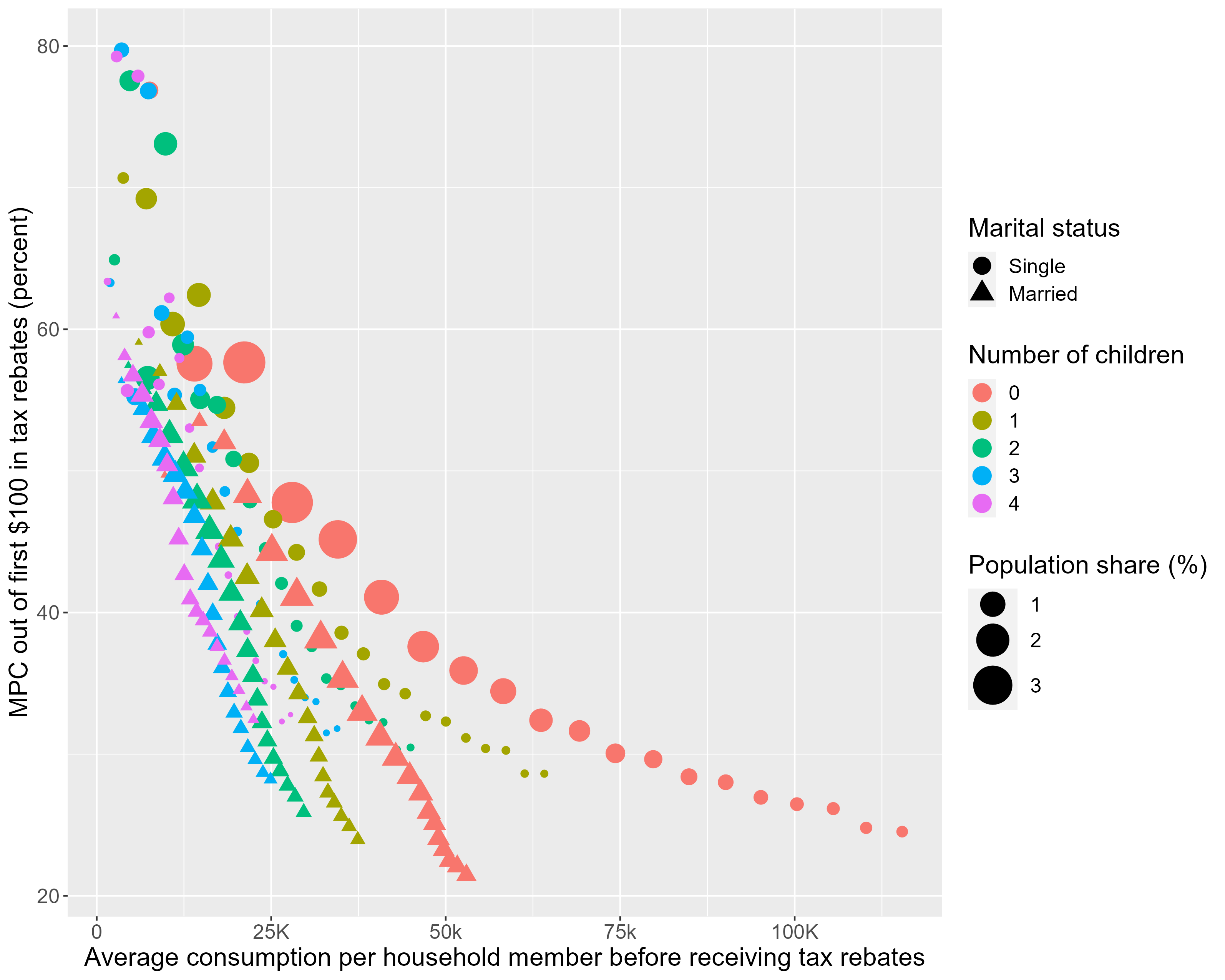}
\smallskip
\caption{Relationship Between Average Consumption Per Household Member Before Receiving Tax Rebates and MPC out of First $\$$100 in Tax Rebates (2008 Calibration).\\
\emph{Notes:} The figure shows the relationship between average consumption per household member before receiving tax rebates and the marginal propensity to consume out of the first $\$100$ in tax rebates for different household types (married vs. single, 0--4 children).}
\label{fig:aveC_vs_MPC}
\end{figure}
\clearpage

\pagebreak
\clearpage

\begin{table}[!t]
\caption[Table]{Dollar Increase in Consumption Following the Receipt of an Additional \$100 Given Initial 2008 Tax Rebate (MPC) and Share of Rebate Consumed Given Transfer Amount (APC)}
\label{table:APCandMPC}
\centering
\begin{threeparttable}
\footnotesize
\begin{tabular}{ll@{\hskip 0.20in}llllll}\toprule\toprule
&   & \multicolumn{5}{c}{MPC given initial transfer amount}\\
\cmidrule(lr){3-7}& Initial cons. (\$) & \$0 &  \$500 &  \$1,000 &  \$2,000 & \$3,000 \\
\toprule
\textbf{Single with 0 children}   \\
Income: 0--20,000 &	13,200	&	60.2	&	59.7	&	53.7	&	48.3	&	48.1	\\
Income: 20,000--40,000 & 	24,500	&	52.8	&	51.0	&	51.0	&	46.9	&	46.5	\\
Income: 40,000--60,000 & 	37,300	&	43.3	&	42.8	&	42.6	&	42.1	&	41.5	\\
\addlinespace[0.5em]
\textbf{Single with 2 children}   \\ 																
Income: 0--20,000 &	4,500	&	75.9	&	57.0	&	57.0	&	57.0	&	57.0	\\
Income: 20,000--40,000 & 	8,500	&	64.6	&	64.6	&	55.7	&	55.5	&	55.5	\\
Income: 40,000--60,000 & 	13,400	&	57.2	&	55.8	&	55.8	&	52.4	&	52.2	\\
\midrule
\textbf{Married with 0 children} 	\\																
Income: 0--20,000 &  	14,500	&	53.3	&	48.8	&	48.7	&	48.7	&	48.7	\\
Income: 20,000--40,000 & 	20,400	&	49.7	&	48.9	&	48.7	&	45.9	&	45.0	\\
Income: 40,000--60,000 & 	27,000	&	42.7	&	42.6	&	42.4	&	40.9	&	39.9	\\
\addlinespace[0.5em]
\textbf{Married with 2 children} 	\\																
Income: 0--20,000 &	 6,600	&	55.9	&	55.9	&	55.9	&	51.0	&	50.8	\\
Income: 20,000--40,000 & 	9,700	&	53.4	&	53.4	&	53.4	&	50.4	&	50.1	\\
Income: 40,000--60,000 &	  13,400	&	49.1	&	49.1	&	48.9	&	48.6	&	47.7	\\
\toprule

&   & \multicolumn{5}{c}{APC given transfer amount}  \\
\cmidrule(lr){3-7}& & \$100 &  \$600 &  \$1,100 &  \$2,100 & \$3,100 \\
\toprule
\textbf{Single with 0 children}   \\
Income: 0--20,000 & &	60.2	&	59.8	&	58.1	&	55.4	&	53.1	\\
Income: 20,000--40,000 &  &	52.8	&	51.7	&	51.4	&	50.5	&	49.3 \\
Income: 40,000--60,000 &  &	43.3	&	43.0	&	42.8	&	42.7	&	42.4	\\
\addlinespace[0.5em]
\textbf{Single with 2 children}   \\ 																
Income: 0--20,000 &  &	75.9	&	64.1	&	60.9	&	59.0	&	58.4	\\
Income: 20,000--40,000 &  &	64.6	&	64.6	&	62.2	&	59.0	&	57.9	\\
Income: 40,000--60,000 &  &	57.2	&	56.3	&	56.0	&	55.6	&	54.5	\\
\midrule

\textbf{Married with 0 children} 	\\																
Income: 0--20,000 &  &	53.3	&	50.4	&	49.7	&	49.2	&	49.1	\\
Income: 20,000--40,000 &  &	49.7	&	49.2	&	49.0	&	48.4	&	47.5	\\
Income: 40,000--60,000 &  &	42.7	&	42.7	&	42.6	&	42.3	&	41.8	\\
\addlinespace[0.5em]
\textbf{Married with 2 children} 	\\																
Income: 0--20,000 &  &	55.9	&	55.9	&	55.9	&	55.3	&	53.9	\\
Income: 20,000--40,000 &  &	53.4	&	53.4	&	53.4	&	52.9	&	52.1	\\
Income: 40,000--60,000 &  &	49.1	&	49.1	&	49.0	&	48.9	&	48.7	\\
\bottomrule
\end{tabular}
\begin{tablenotes}[flushleft]
\footnotesize
\item \textit{Notes}: The first part of the table reports the marginal propensity to consume (MPC) out of an additional $\$$100 in the size of a potential tax rebate given initial transfer amounts for different household types (single or married, 0 or 2 children, and different household incomes). That is, it reports the value of $MPC^{\$100}(g; D) = \frac{\Delta c(g)}{(D+\$100)-D}$, where $\Delta c(g)$ is the change in consumption of a household of type $g$ whose transfer increases from $D\geq0$ to $D+\$100$. The column titled ``Initial cons.'' shows average consumption per household member given household type if the household does not receive tax rebates. The second part of the table reports the average propensity to consume (APC) given transfer amount for the same household types. That is, it reports the value of $APC(g; D) = \frac{\Delta c(g)}{D}$, where $\Delta c(g)$ is the change in consumption of a household of type $g$ who receives a transfer of $D>0$.
\end{tablenotes}
\end{threeparttable}
\end{table}
\clearpage
\begin{table}[!t]
\caption[Table]{Resource Equivalent Variation (REV): Budget Savings Relative to the Actual Policy Given Allocation Constraint}\label{table:REV}
\centering
\begin{threeparttable}
\small
\begin{tabular}{ll@{\hskip 0.60in}c}\toprule\toprule\\
&  Allocation Constraints & REV  \\ \toprule \\ \multicolumn{2}{l}{\textbf{Vary income-eligibility criteria}} \\
(1) & Adjust income-specific eligibility criteria &	2.0   \\ \midrule \\
\multicolumn{2}{l}{\textbf{Vary income-eligibility criteria and max. check amount}} \\
(2) & Maximum limit per adult: $\$$900 &	8.8   \\
(3) & Maximum limit per adult: $\$$1,200 &	11.4	   \\
(4) & Maximum limit per child: $\$$600 &	12.0	   \\
(5) & Maximum limit per child: $\$$900  &	15.5	 \\
(6) & Maximum limit: $\$$900 per adult and $\$$600 per child &	13.9  \\
(7) & Maximum limit: $\$$1,200 per adult and $\$$900 per child&	16.9 \\
\bottomrule
\end{tabular}
\footnotesize
\begin{tablenotes}[flushleft]
\footnotesize
\item \textit{Notes}: The REV numbers specify the percentage by which the government can reduce total spending on tax rebates and still achieve the same stimulus as the actual policy. Rows correspond to different allocation constraints. A REV value of 0 percent means the government cannot reduce spending at all if it wants to maintain the same increase in aggregate consumption as under the actual policy, whereas a value of 99 percent means the government can reduce total spending by 99 percent and still achieve the same increase in aggregate consumption as under the actual policy. The actual 2008 policy was income-tested and had a maximum tax rebate amount of $\$$600 per adult and $\$$300 per child.
\end{tablenotes}
\end{threeparttable}
\end{table}

\clearpage
\pagebreak

\appendix

\setlength{\footnotemargin}{5.0mm}
\begingroup
\doublespacing
\centering
\Large ONLINE APPENDIX \\
\Large\begin{singlespace}\href{\PAPERDOIURL}{\PAPERTITLE}\end{singlespace}
\large \AUTHORNYGAARD, \AUTHORSORENSEN, and \AUTHORWANG\\[1.0em]
\endgroup

\renewcommand{\thefigure}{A.\arabic{figure}}
\setcounter{figure}{0}
\renewcommand{\thetable}{A.\arabic{table}}
\setcounter{table}{0}
\renewcommand{\theequation}{A.\arabic{equation}}
\setcounter{equation}{0}
\renewcommand{\thefootnote}{A.\arabic{footnote}}
\setcounter{footnote}{0}

\section{Calibration}
\label{sec:Appendix_Calibration}

This section provides additional details about the calibration of the model. A summary of the parameters that are determined outside the equilibrium is reported in Table~\ref{Table:Non_calibrated_parameters}. Table~\ref{Table:Calibrated_parameters} provides a corresponding summary of the parameters that are determined jointly in equilibrium.

\textit{Life-cycle parameters}---We use life-tables for the U.S. Social Security Area (SSA) for the year 2020 to derive age-specific survival probabilities, $\psi_{j}$. Life-tables reported by the SSA are gender-specific. We obtain non-gender-specific survival probabilities by combining the age- and gender-specific survival probabilities from the SSA with data on the distribution of gender by age reported by the Census. We normalize the mass of 18-year-olds in the model to 1 and let the size of new cohorts increase at 1.1 percent per year to match data from the Census.

We use data from the Panel Study of Income Dynamics (PSID) for the period 1997--2017 to obtain the probability of being college educated, the probability of being married, and the initial distribution of children. We assume that 30.3 percent of agents are college-educated, which corresponds to the share of 18+ year-old household heads in the PSID with at least a bachelor's degree or a minimum of 4 years of college education. We assume that 43.7 percent of non-college-educated agents and 56.4 percent of college-educated agents are married, which corresponds to the share of 18+ year-old  household heads in the PSID that are married by college attainment. Finally, we let the initial distribution of children be equal to the distribution of children under the age of 18 for 18--25 year-old household heads in the PSID, with the number of children top-coded at 4. We condition the initial distribution of children on the household head's college attainment and marital status. Doing so allows us to account for the observation that young college-educated individuals are less likely to have children than young non-college-educated individuals and that young married individuals are more likely to have children than young non-married individuals. Table~\ref{Table:Initial_family_composition_conditions} summarizes the initial conditions for college attainment, marital status, and number of children.

\textit{Transition probabilities for number of children}---The PSID has been administered biennially since 1997. Because a period in the model is one year, we use data from the PSID for the period 1993--1997 to derive transition probabilities for the number of children under the age of 18. We do this by estimating an ordered logistic regression of the number of children under the age of 18 at time $t+1$ conditional on the household head's age, age squared, marital status, college attainment, and number of children under the age of 18 at time $t$. The regression results are reported in Table~\ref{Table:Transition_probabilities_regression}. The transition probabilities for the number of children under the age of 18 are then given by the standard ordered logistic formula:
\begin{equation}
\mathbb{P}\left(k^{\prime}=i\vert \boldsymbol{x}\right)=\frac{1}{1+\exp(-\tau_i+\boldsymbol{x}\bm{\beta})}-\frac{1}{1+\exp(-\tau_{i-1}+\boldsymbol{x}\bm{\beta})} \, ,
\end{equation}
where $\boldsymbol{x}=\left(k,j,j^{2},m,e\right)$ is a vector with the number of children ($k$) under the age of 18 at time $t$ ($k'$ is the number of children under age of 18 at time $t+1$), the age ($j$) and age-squared ($j^2$) of the agent, the marital status ($m$) of the agent, and the educational attainment ($e$) of the agent. Similarly, $\bm{\beta}$ is a vector of parameters and $\boldsymbol{\tau}$ is a vector of constants of increasing size which can be interpreted as hurdles for the ordered outcomes.

\textit{Spousal income}---We use data for married individuals in the PSID for the period 1997--2017 to estimate spousal income. To do this, we first estimate the following OLS regression:
\begin{equation}
\label{eq:Spousal-income}
\begin{array}{llr}
\multirow{2}{*}{$\ln(y_{t}^{S})=\Bigg\{$}&\beta_{0}+\beta_{1}\ln(y_{t}^{H})+\sum_{p=2}^{4}\beta_{p}j_{t}^{p-1}+\beta_{5}\mathbbm{1}_{e_{t}=1}+\sum_{q=1}^{4}\gamma_{q}\mathbbm{1}_{k_{t}=q} & j<j_{R}\\
&\beta_{0}+\beta_{1}\ln(y_{t}^{H})+\sum_{p=2}^{4}\beta_{p}j_{t}^{p-1}+\beta_{5}\mathbbm{1}_{e_{t}=1}+\sum_{q=1}^{4}\gamma_{q}\mathbbm{1}_{k_{t}=q}+\beta_{6}k_{t}+\beta_{7} & j\geq j_{R},
\end{array}
\end{equation}
where $\ln(y_{t}^{S})$ and $\ln(y_{t}^{H})$ denote the logarithm of the spouse's and reference person's income, respectively, with income defined as the sum of labor earnings, Social Security benefits, Supplemental Security Income, UI benefits, and other transfers. Next, $\mathbbm{1}_{e_{t}=1}$ is an indicator function that is equal to one if the reference person is college-educated and $\mathbbm{1}_{k_{t}=q}$ is an indicator function that is equal to one if the reference person has $q\in\left\{1,2,3,4\right\}$ children under the age of 18 at time $t$. Lastly, $\beta_{6}$ enables us to capture that the ``child penalty'' (i.e., the reduction in average spousal income due to children) is different for young and old individuals, and $\beta_{7}$ allows for differences in the intercept term for young and old individuals.\footnote{We do not use any information about the spouse, such as the spouse's age or educational attainment, in our regressions because we do not keep track of the spouse's idiosyncratic state in the model.} We abbreviate from Equation~\eqref{eq:Spousal-income} year fixed-effects and error terms that are present in the regression. The regression results are reported in Table~\ref{Table:Spousal_income_regression}. We use this regression to predict average spousal income conditional on the idiosyncratic state of the household head and use the variance of the residual from this regression to obtain an estimate of the variance of spousal income. Finally, we discretize both the household head's and the spouse's income process by means of the Tauchen method. Households draw their initial productivity shocks from the stationary distribution of the household head's and the spouse's income process.

The benchmark analysis studied in the main text assumes that spousal income shocks are i.i.d.\footnote{The only exception is during times of unemployment. In particular, for computational simplicity we assume that the unemployment shock for the head and the spouse (both during the 2008--2009 and 2020--2021 crises) are perfectly correlated. In the event of unemployment, spousal income is given by $B_{j,e,k,\eta,\nu,\xi,b}=B_{j,e,k,\eta,\nu}[\xi+b(1-\xi)]$, where $B_{j,e,k,\eta,\nu}$ is as defined in Equation \eqref{eq:Spousal-income}.} This leads to a correlation between the logarithm of the spouse's and reference person's income of 0.22, compared to a corresponding correlation of 0.26 in the PSID. The optimal allocation results are robust to allowing for persistent spousal productivity shocks.
\clearpage 

\begin{table}[!t] \caption[Table]{Parameters Determined Outside the Model}\label{Table:Non_calibrated_parameters}      \centering      \begin{threeparttable}        \scriptsize        \begin{tabular}{l@{\hskip 0.2in}llr}\toprule\toprule Parameter & Description & Source & Value \\ \toprule \\ \multicolumn{2}{l}{\textbf{Life-cycle parameters}} \\ $J$ & Maximum lifespan = 100 & & 83 \\ $j_R$ & Retirement age = 65 & & 48 \\ $\gamma$ & Risk aversion & IES = 0.5 & 2.000 \\ $\psi_j$ & Age-specific survival probabilities & SSA life-tables & \\  & Growth rate of 18-year-olds & Census & 0.011 \\ $\mathbb{P}_{k^{\prime}\vert(j,e,m,k)}$ & Transition probabilities for number of children & PSID & \\ $\delta$ & Discount factor, $\delta\in\{\underline{\delta},\overline{\delta}\}$ & & \{0.600,0.950\} \\ $\mathbb{P}_{\delta=\underline{\delta}\vert e=0}$ & Share of non-college-educated with $\delta=\underline{\delta}$ & \textcite{jappelliIntertemporalChoiceConsumption2006} & 0.400 \\ $\mathbb{P}_{\delta=\underline{\delta}\vert e=1}$ & Share of college-educated with $\delta=\underline{\delta}$ & \textcite{jappelliIntertemporalChoiceConsumption2006} & 0.100 \\ \\ \multicolumn{2}{l}{\textbf{Technology and income parameters}} \\ $r$ & Real interest rate & \textcite{mcgrattanAverageDebtEquity2003} & 0.040 \\ $f_{j,e}$ & Age- and educ.-specific deterministic labor prod. & \textcite{conesaImplicationsIncreasingCollege2020} & \\ $\rho$ & Persistence of ref. person's AR(1) shocks & \textcite{pashchenkoQuantitativeAnalysisHealth2013} &  0.980 \\ $\sigma_\mu^2$ & Variance of ref. person's AR(1) shocks & \textcite{pashchenkoQuantitativeAnalysisHealth2013} & 0.018 \\ $B_{j,e,k,\eta,\nu}$ & Spousal income if married & PSID & \\ \\ \multicolumn{2}{l}{\textbf{Taxation}} \\ $a_0$ & Tax parameter & \textcite{gouveiaEffectiveFederalIndividual1994} & 0.258 \\ $a_1$ & Tax parameter & \textcite{gouveiaEffectiveFederalIndividual1994} & 0.768 \\ \\ \multicolumn{2}{l}{\textbf{2008--2009 crisis (Great Recession)}} \\ $1-\xi$ & Unemployment duration (percent of year) & BLS  & 0.468 \\ $\pi^{U}(j,e)$ & Age- and education-specific unemp. prob. & BLS \\ \\ \multicolumn{2}{l}{\textbf{2020--2021 crisis (COVID-19 pandemic)}} \\ $1-\xi$ & Unemployment duration (percent of year) & BLS  & 0.349 \\ $\pi^{U}(j,e,\eta)$ & Age- and earnings-specific unemp. prob. & \textcite{cajnerLaborMarketBeginning2020} \\ \bottomrule        \end{tabular}        \begin{tablenotes}[flushleft]          \scriptsize          \item \textit{Notes}: The table lists the parameters that are determined outside the model. See the text for details.  \end{tablenotes}      \end{threeparttable}    \end{table}

\clearpage

\begin{table}[!t] \caption[Table]{Parameters Determined Jointly in Equilibrium}\label{Table:Calibrated_parameters}      \centering      \begin{threeparttable}        \scriptsize        \begin{tabular}{l@{\hskip 0.2in}lllr}\toprule\toprule Parameter & Description & Value & Target & Model \\ \toprule \\ \multicolumn{2}{l}{\textbf{Non-crisis-specific parameters }} \\ $\theta$ & Normalization of model units & 0.565 & Median household income = 1.000 & 1.000 \\ $SS_{e=0}$ & Social Security non-college-educated & 0.220 & Avg. SS non-college/median HH inc. = 0.220 & 0.220 \\ $SS_{e=1}$ & Social Security college-educated & 0.266 & Avg. SS college/avg. SS non-college = 1.209 & 1.209 \\ $\gamma$ & Government consumption/GDP & 0.176 & Government cons./GDP = 0.176 & 0.176 \\ \\ \multicolumn{2}{l}{\textbf{2008--2009 crisis (Great Recession)}} \\ $b$ & UI replacement rate & 0.375 & Total UI benefits/total wages = 0.021 & 0.021  \\ \\ \multicolumn{2}{l}{\textbf{2020--2021 crisis (COVID-19 pandemic)}} \\ $\kappa$ & Drop in marginal utility of cons. & 0.669 & Percent drop in agg. cons. due to lockdown = 0.109 & 0.109  \\ \bottomrule        \end{tabular}        \begin{tablenotes}[flushleft]          \scriptsize          \item \textit{Notes}: The table lists the parameters that are determined jointly in equilibrium. Numbers in the model are normalized such that a value of 1.0 in the steady state prior to the 2008--2009 crisis corresponds to $\$$54,831, whereas a value of 1.0 in the steady state prior to the 2020--2021 crisis corresponds to $\$$62,502 (in 2012 USD). Social Security benefits are calibrated to match data on Social Security benefits for 65+ year-olds with positive Social Security benefits in the CPS. The level of government consumption is equal to $G=\gamma Y$, where $Y$ is GDP.   \end{tablenotes}      \end{threeparttable}    \end{table}

\clearpage

\begin{table}[!t] \caption[Table]{Initial Conditions for College Attainment, Marital Status, and Number of Children}\label{Table:Initial_family_composition_conditions}      \centering      \begin{threeparttable}        \footnotesize        \begin{tabular}{l@{\hskip 0.7in}lr}\toprule\toprule Description & Source & Value \\ \toprule \\ \multicolumn{2}{l}{\textbf{College attainment and marital status}} \\ 
Probability of being college-educated & PSID & 0.303 \\ 
Probability of being single for non-college-educated & PSID & 0.564 \\ Probability of being single for college-educated & PSID & 0.436  \\ \\
\multicolumn{2}{l}{\textbf{Number of children for non-college-educated and single}} \\ 
0 children & PSID & 0.733 \\ 1 child & PSID & 0.151 \\ 2 children & PSID & 0.083 \\ 3 children & PSID & 0.024 \\ 4 children & PSID & 0.009 \\ \\ \multicolumn{2}{l}{\textbf{Number of children for college-educated and single}} \\ 
0 children & PSID & 0.975 \\ 1 child & PSID & 0.024 \\ 2 children & PSID & 1E-04 \\ 3 children & PSID & 0.001  \\ 4 children & PSID & 0.000 \\ \\ 
\multicolumn{2}{l}{\textbf{Number of children for non-college-educated and married}} \\ 
0 children & PSID & 0.414  \\ 1 child & PSID & 0.296  \\ 2 children & PSID & 0.213  \\ 3 children & PSID & 0.057  \\ 4 children & PSID & 0.020  \\ \\ \multicolumn{2}{l}{\textbf{Number of children for college-educated and married}} \\ 
0 children & PSID & 0.753  \\ 1 child & PSID & 0.215   \\ 2 children & PSID &  0.022 \\ 3 children & PSID & 0.009  \\ 4 children & PSID & 0.000 \\ \bottomrule        \end{tabular}        \begin{tablenotes}[flushleft]          \footnotesize          \item \textit{Notes}: College attainment and marital status are given by the share of 18+ year-old household heads in the PSID that are married and college-educated, where a college-degree is defined as having at least a bachelor's degree or a minimum of 4 years of college. The initial distribution of children is given by the distribution of children under the age of 18 for 18--25 year-old household heads by marital status and college attainment in the PSID.   \end{tablenotes}      \end{threeparttable}    \end{table}

\clearpage

\begin{table}[!t] \caption[Table]{Transition Probabilities for Number of Children: Ordered Logistic Regression Results}\label{Table:Transition_probabilities_regression}      \centering      \begin{threeparttable}        \footnotesize        \begin{tabular}{l@{\hskip 1.0in}c}\toprule\toprule Dependent variable: Number of children at $t+1$: $k_{t+1}\in\{0,\dots,4\}$  &  \\ \toprule \\ Reference person has 1 child at $t$: $\mathbb{I}(k_t=1)$  & 4.970 \\ & (0.093) \\ Reference person has 2 children at $t$: $\mathbb{I}(k_t=2)$  & 8.720 \\ & (0.147) \\ Reference person has 3 children at $t$: $\mathbb{I}(k_t=3)$  & 12.969 \\ & (0.215) \\ Reference person has 4 children at $t$: $\mathbb{I}(k_t=4)$  &  17.569 \\ & (0.338) \\ Age of reference person & -0.082 \\ & (0.011) \\ Age squared of reference person & 3E-04 \\ & (1E-04) \\ Marital status of reference person: $\mathbb{I}(m_t=1)$ & 0.387 \\ & (0.052) \\ College attainment of reference person: $\mathbb{I}(e_t=1)$ & 0.134 \\ & (0.047) \\ \midrule Cut 1 & 0.659 \\ & (0.215) \\ Cut 2 & 4.479 \\ & (0.210) \\ Cut 3 & 9.005 \\ & (0.228) \\ Cut 4 & 13.359 \\ & (0.286) \\ \\ Pseudo $R^2$ & 0.6912 \\ Number of observations & 27,660 \\  \bottomrule        \end{tabular}        \begin{tablenotes}[flushleft]          \footnotesize          \item \textit{Notes}: The table reports results from an ordered logistic regression of the household head's number of children under the age of 18 at time $t+1$ on the household head's number of children under the age of 18 at time $t$, a quadratic in the household head's age at time $t$, the marital status of the household head at time $t$, and the college attainment of the household head at time $t$. Number of children has been top-coded at 4. Standard errors are reported in parentheses. Data source: PSID.  \end{tablenotes}      \end{threeparttable}    \end{table}

\clearpage

\begin{table}[!t] \caption[Table]{Spousal Income: Ordinary Least Squares Regression Results}\label{Table:Spousal_income_regression}      \centering      \begin{threeparttable}        \footnotesize        \begin{tabular}{l@{\hskip 1.0in}c}\toprule\toprule Dependent variable: Logarithm of spousal income  &  \\ \toprule \\ Logarithm of reference person's income & 0.134 \\ & (0.012) \\ Age of reference person & 0.164 \\ & (0.014) \\ Age squared of reference person & -0.003 \\ & (3E-04) \\ Age cubed of reference person & 1E-05 \\ & (2E-06) \\  College attainment of reference person: $\mathbb{I}(e=1)$ & 0.206 \\ & (0.016) \\  Reference person has 1 child: $\mathbb{I}(k=1)$ & -0.168 \\ & (0.020) \\ Reference person has 2 children: $\mathbb{I}(k=2)$ & -0.302 \\ & (0.021) \\ Reference person has 3 children: $\mathbb{I}(k=3)$ & -0.462 \\ & (0.033) \\ Reference person has 4 children: $\mathbb{I}(k=4)$ & -0.786 \\ & (0.058)  \\ Reference person is 65+: $\mathbb{I}(j\geq j_{R})$ & -0.223 \\ & (0.039) \\ Interaction between 65+ dummy and number of children: $\mathbb{I}(j\geq j_{R})k$ & 0.173 \\ & (0.059)  \\ Constant term & -3.407 \\ & (0.217)  \\ \\ $R^2$ & 0.1405 \\ Number of observations & 30,410  \\ \bottomrule        \end{tabular}        \begin{tablenotes}[flushleft]          \footnotesize          \item \textit{Notes}: The table reports results from an ordinary least squares regression of the logarithm of spousal income on the logarithm of the household head's income, a cubic in the age of the household head, the educational attainment of the household head, the household head's number of children under the age of 18, a dummy variable for whether or not the household head is at least 65 years old, and an interaction term between the 65+ dummy variable and the number of children of the household head. The regression also includes year fixed-effects. Income is given by the sum of labor earnings, Social Security benefits, Supplemental Security Income, UI benefits, and other transfers. Number of children has been top-coded at 4. The sample is restricted to married household heads. Standard errors are reported in parentheses. Data source: PSID.  \end{tablenotes}      \end{threeparttable}    \end{table}

\clearpage

\renewcommand{\thefigure}{B.\arabic{figure}}
\setcounter{figure}{0}
\renewcommand{\thetable}{B.\arabic{table}}
\setcounter{table}{0}
\renewcommand{\theequation}{B.\arabic{equation}}
\setcounter{equation}{0}
\renewcommand{\thefootnote}{B.\arabic{footnote}}
\setcounter{footnote}{0}

\section{Additional results}
\label{sec:Appendix_additional_results}

\begin{figure}
\centering
\caption{Optimal Allocation Queue}\label{fig:queue_illustration}
\includegraphics[height=12cm, width=15cm]{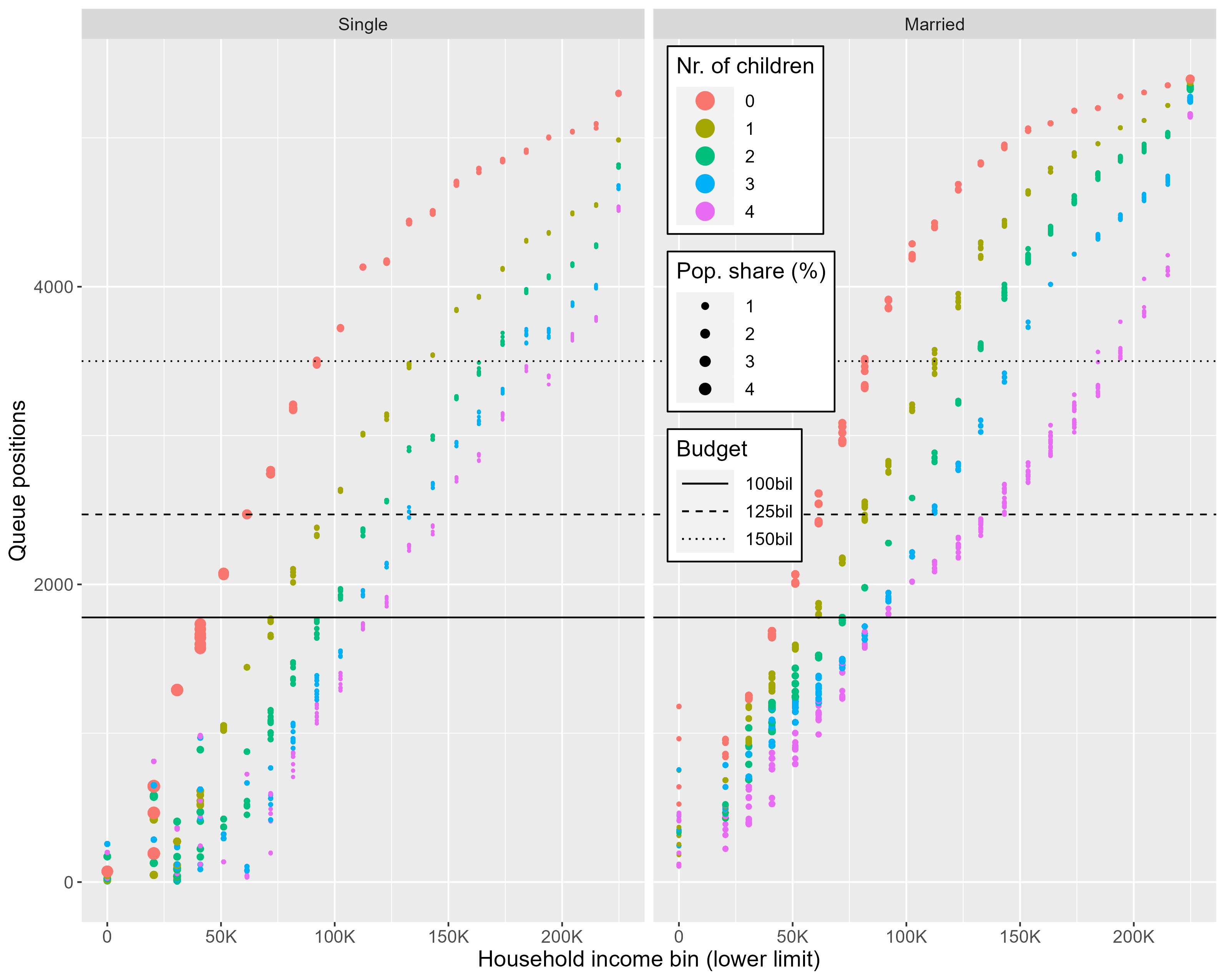}
\begin{minipage}{\linewidth}
\vspace*{4mm}
\footnotesize \emph{Notes:} The graph plots the optimal consumption stimulus allocation queue for the case with a maximum tax rebate of \$900 per adult and \$600 per child. The left panel plots the queue position for single households with 0--4 children and different income bins, and the right panel plots the queue position for corresponding married households. While the total number of queue positions (Y-axis) that will receive allocations is governed by the overall budget, the ranking of allocation increments is invariant to aggregate resources.
\end{minipage}\end{figure}

\textit{Illustration of the optimal allocation queue}--Figure~\ref{fig:queue_illustration} illustrates the optimal consumption stimulus allocation queue for the case with a maximum tax rebate of \$900 per adult and \$600 per child. The left-hand panel plots the queue positions for single households with different income levels and 0--4 children, and the right-hand panel plots the queue positions for corresponding married households. The Y-axis marks out the $Q_{g,l}$ queue positions defined in Equation~\eqref{eq:optimal_allocation_queue}, where lower numbered (higher-ranked) queue positions receive allocations first as aggregate resources increase. The multiple markers for the same household type follow from the assumption that the planner allocates tax rebates in \$100 increments and that the same household can receive multiple rebates. While the queue is invariant to the total budget available for tax rebates, the budget still governs what allocations are feasible. In Figure~\ref{fig:queue_illustration}, dots below the budget line indicate increments that are below the cut-off to be funded. In our figure, we illustrate what policies are chosen with a \$100, \$125, and \$150 billion budget.

\begin{figure}
    \centering
   \caption{Actual vs. Optimal Allocation by Income, Family Status, and Planner Consumption Inequality Aversion. Maximum Tax Rebate $\$$900 per adult and $\$$600 per child}\label{fig:2008-policy-different-lambda}
    \includegraphics[height=12cm, width=15cm]{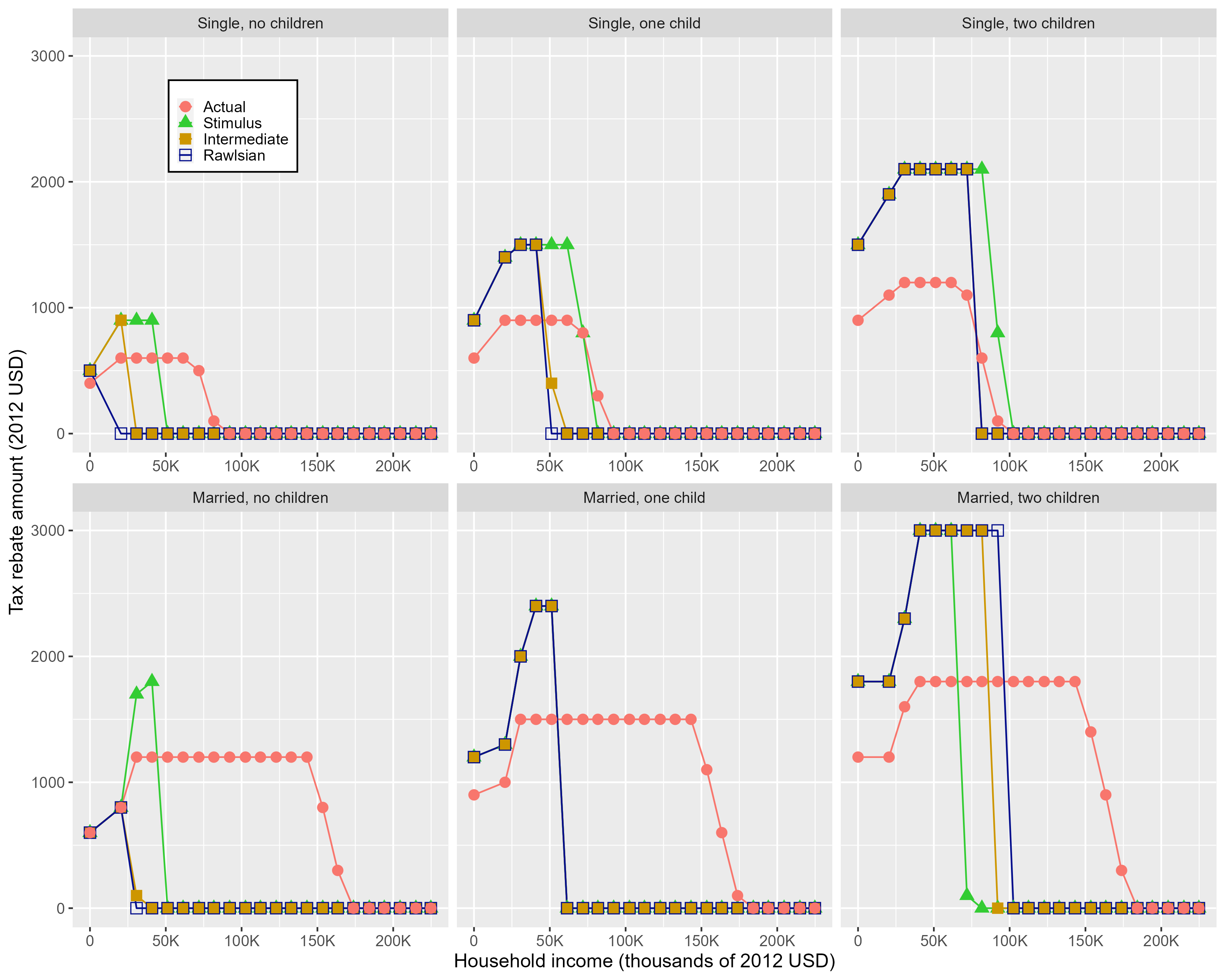}
\begin{minipage}{\linewidth}
\vspace*{4mm}
\footnotesize \emph{Notes:} The figure shows the allocation of tax rebates by household income and family status (marital status and number of children) for the 2008 policy and for the optimal allocation of a planner with very high (Rawlsian, $\lambda=-99$), intermediate ($\lambda=-1$), and no aversion to consumption inequality ($\lambda=1$) of the same amount of money calculated under the assumption that the maximum tax rebate is $\$900$ per adult and $\$600$ per child.
\end{minipage}\end{figure}

\textit{Optimal allocation for planners with different consumption inequality aversion}---In Section~\ref{sec:Results-2008}, we compared the actual tax rebates in 2008 with the optimal policy assuming the government's objective was to maximize the stimulus effect on consumption in that year. Figure~\ref{fig:2008-policy-different-lambda} studies the sensitivity of the optimal policy to different degrees of planner 2008 consumption inequality aversion.\footnote{The planner's optimization problem when the government exhibits aversion to 2008 consumption inequality is defined analogously to Equation~\eqref{eq:Planner_objective_emp_app-2021}.} In particular, we plot the allocation of tax rebates by household income and family status (marital status and number of children) for the 2008 policy and for the optimal allocation of a planner with very high (Rawlsian, $\lambda=-99$), intermediate ($\lambda=-1$), and no aversion to consumption inequality ($\lambda=1$) of the same amount of money calculated under the assumption that the maximum tax rebate is $\$900$ per adult and $\$600$ per child. The optimal stimulus planner studied in Section~\ref{sec:Results-2008} corresponds to the planner with $\lambda=1$. We find that the optimal policy is largely robust to different degrees of planner consumption inequality aversion. This follows because low-consumption households have higher MPCs in the model due to concave utility, as illustrated in Figure~\ref{fig:aveC_vs_MPC}. Both limit-planners thus optimally prioritize low-consumption households when deciding how to allocate the tax rebates.

\begin{figure}
    \centering
    \caption{Sensitivity Analysis: Optimal Welfare Allocation by Income and Family Status for Planner with Intermediate Inequality Aversion. Maximum Check Size $\$$2,000 per adult and $\$$2,000 per child}\label{fig:sensitivity}
    \begin{subfigure}[t]{\linewidth}
    \caption{Vary the share of impatient households}
  \begin{tabular}{@{\hskip-0.25in}cccc@{}}
    \includegraphics[width=.29\textwidth]{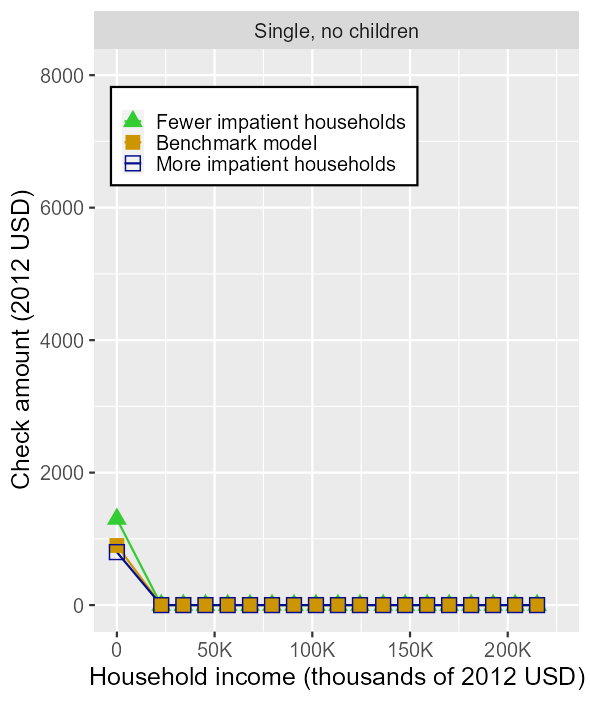} &
    \includegraphics[width=.252\textwidth]{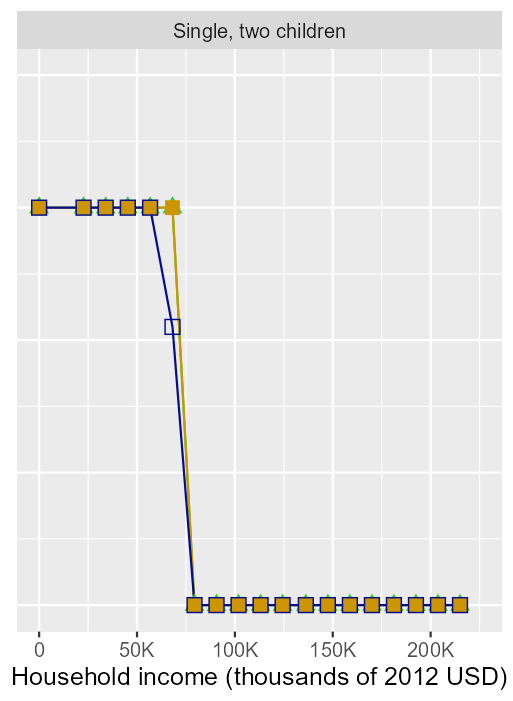} &
    \includegraphics[width=.252\textwidth]{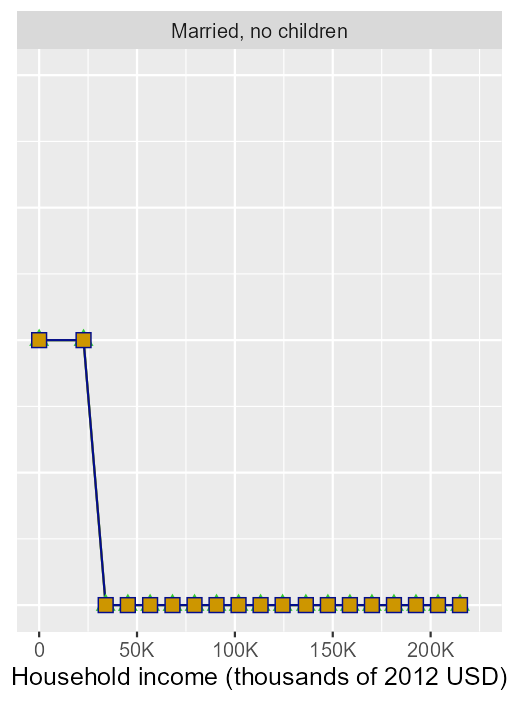} &
    \includegraphics[width=.252\textwidth]{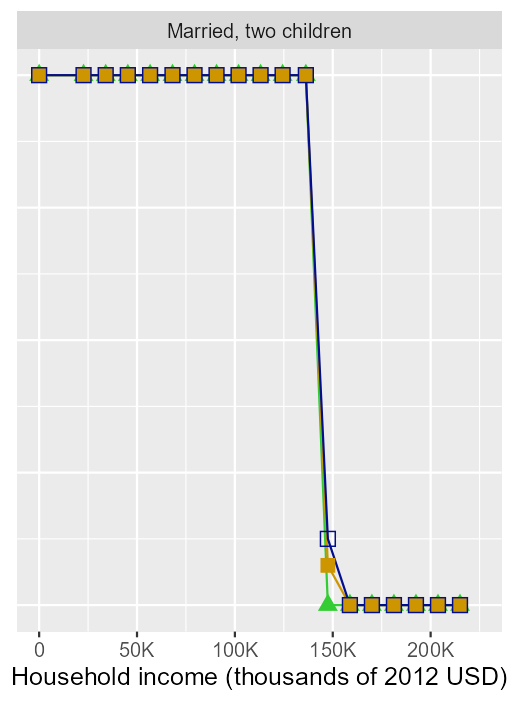}
   \end{tabular}
     \vspace{0.3cm}
    \end{subfigure}

    \begin{subfigure}[t]{\linewidth}
    \caption{With and without unemployment insurance (UI) benefits}
  \begin{tabular}{@{\hskip-0.25in}cccc@{}}
    \includegraphics[width=.29\textwidth]{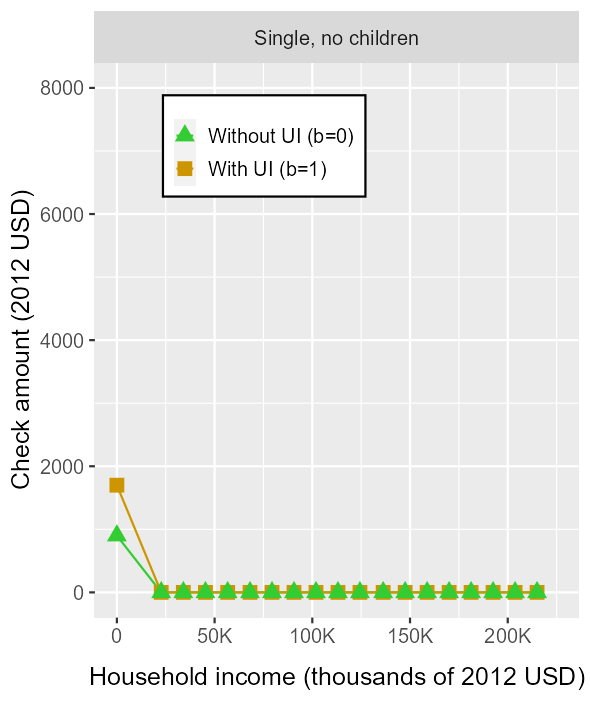} &
    \includegraphics[width=.252\textwidth]{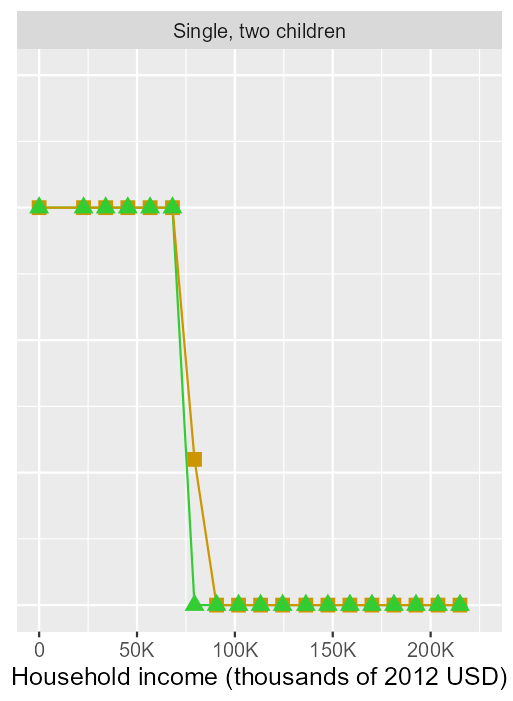} &
    \includegraphics[width=.252\textwidth]{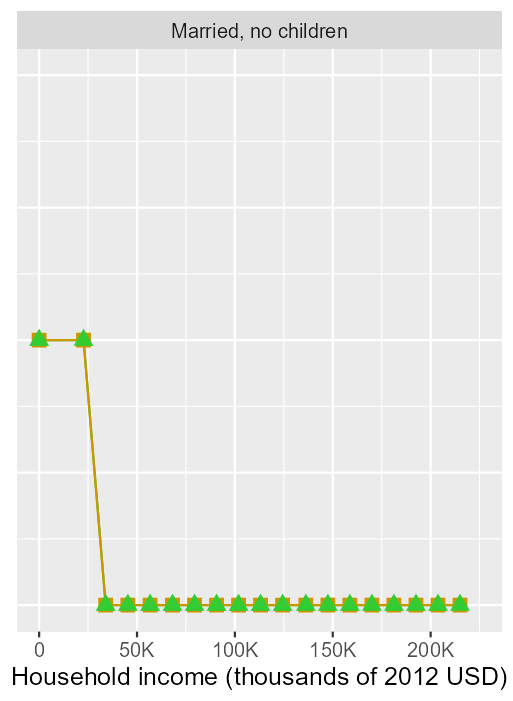} &
    \includegraphics[width=.252\textwidth]{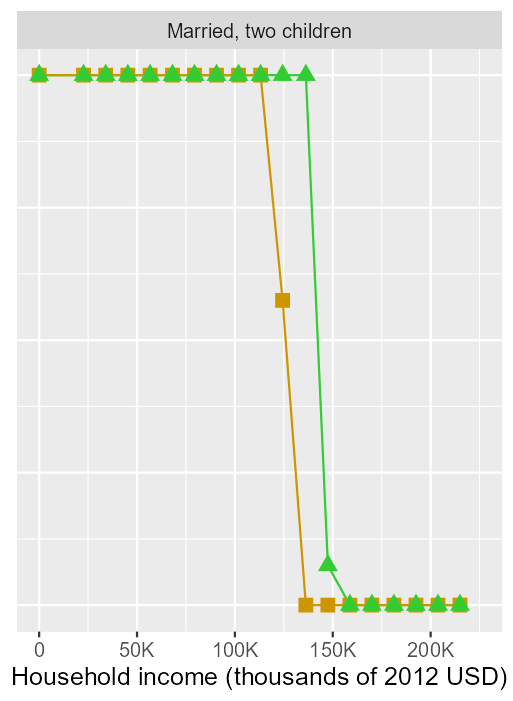}
   \end{tabular}
     \vspace{0.3cm}
    \end{subfigure}

    \begin{subfigure}[t]{\linewidth}
    \caption{Vary the real interest rate}
  \begin{tabular}{@{\hskip-0.25in}cccc@{}}
    \includegraphics[width=.29\textwidth]{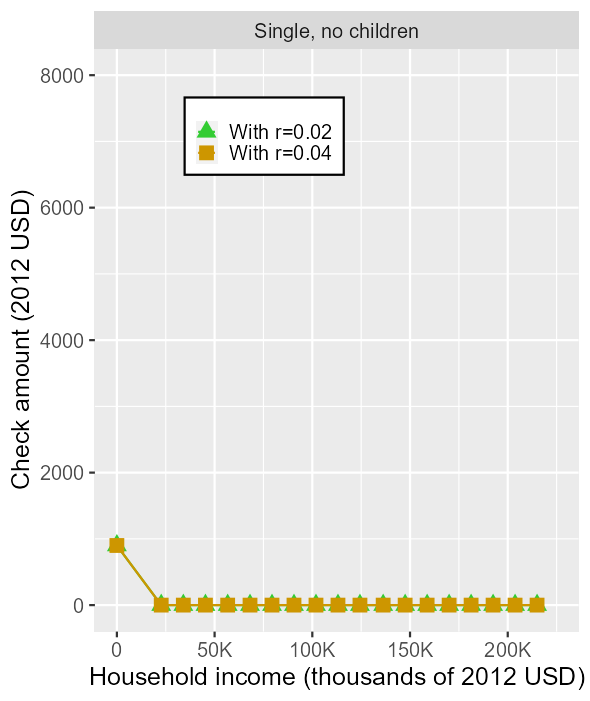} &
    \includegraphics[width=.252\textwidth]{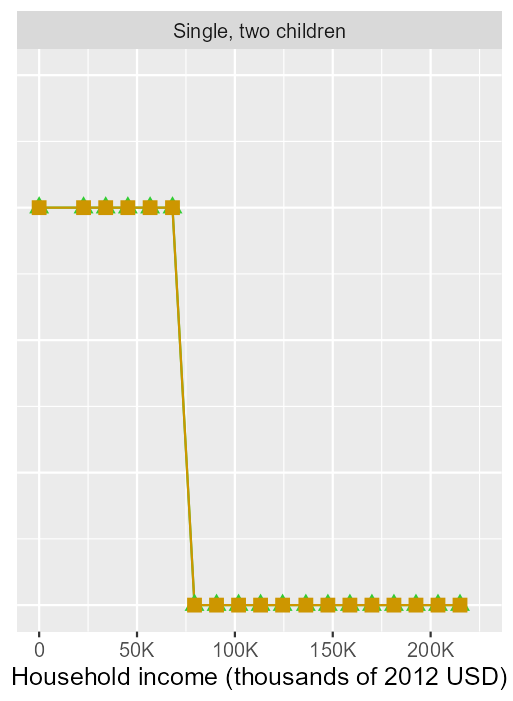} &
    \includegraphics[width=.252\textwidth]{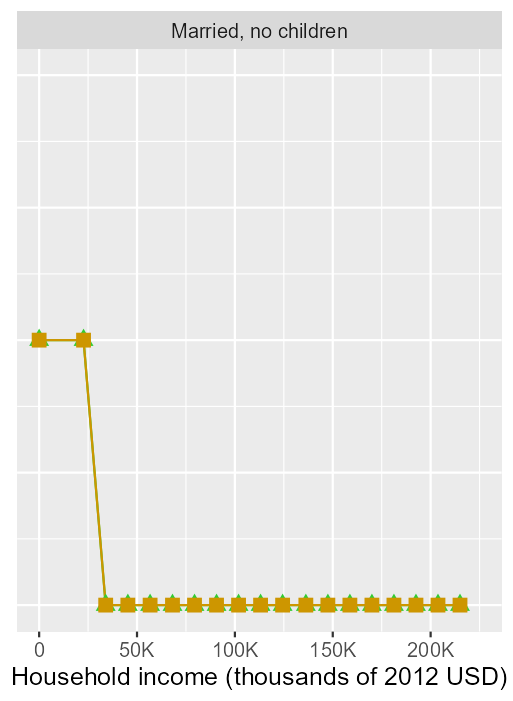} &
    \includegraphics[width=.252\textwidth]{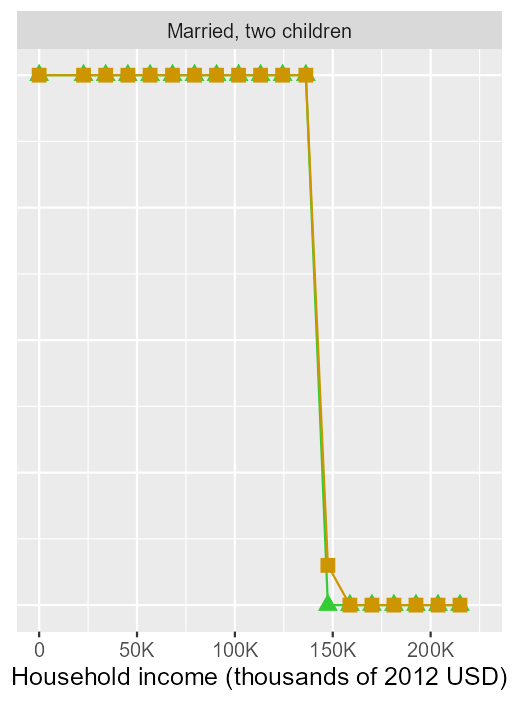}
   \end{tabular}
    \end{subfigure}
\begin{minipage}{\linewidth}
\vspace*{4mm}
\footnotesize \emph{Notes:} The graphs show the sensitivity of the optimal welfare allocation results for the 2021 policy---by income and family status for the case with intermediate planner inequality version ($\lambda=-1$) and a maximum check limit of $\$$2,000 per adult and child---to alternative model parameterizations. The top panel shows the results when we vary the share of impatient households (for whom the the discount factor $\delta$ equals the lower value $\underline{\delta}$). Fewer (more) impatient households refer to the model where we reduce (increase) the mass of impatient households by 50 percent. The middle panel compares the results from the benchmark model with UI benefits and the model without UI benefits. The bottom panel compares the results when the real interest rate is 2 percent rather than 4 percent as in the benchmark model.
\end{minipage}
\end{figure}

\textit{Sensitivity analysis}---Figure~\ref{fig:sensitivity} explores if the optimal welfare allocations in 2021 for the intermediate planner ($\lambda=-1$), facing the same individual check limits and overall budget as described in Section~\ref{sec:Results-2021}, are robust to various modeling choices. To keep the number of graphs manageable, we only show the results for singles and married with either 0 or 2 children. First, there is large variation in MPC estimates in the literature. Given that the MPC estimates are sensitive to the choice of discount factor heterogeneity in the model and that this might have implications for the optimal allocation of checks, we start by studying the sensitivity of the optimal allocations to increasing and reducing the mass of impatient (i.e., $\delta=\underline{\delta}$) households by 50 percent.
As shown in panel~(a) of Figure~\ref{fig:sensitivity}, the optimal allocations are robust to varying the mass of impatient households.

Second, motivated by the increased UI replacement rate during the pandemic, the benchmark analysis for the 2020--2021 crisis assumes that UI replaces 100 percent of lost earnings. The increased generosity of UI benefits may affect labor-supply, which is an issue that we do not address in this paper. Instead, we ask the question if the optimal allocation of checks depends on the level of UI benefits. Panel~(b) shows that the allocations are highly robust to two extreme choices for UI benefits when we assume (for consistency with the actual policy) that the planner allocates the checks conditional on the households' last year's income: no UI benefits ($b=0$) and complete UI benefits ($b=1$).\footnote{The finding that the optimal allocations are not sensitive to the choice of UI benefits is robust to alternative durations of the unemployment spell.}

Third, the interest rate in the benchmark model is set to 4 percent as in \textcite{mcgrattanAverageDebtEquity2003}. The interest rate affects the propensity to save out of the checks. We test the sensitivity of our optimal allocations by reducing the interest rate to 2 percent. As shown in panel~(c), the allocations are robust to alternative values for the interest rate.
\clearpage 

\clearpage

\clearpage

\clearpage
\pagebreak

\renewcommand{\thefigure}{C.\arabic{figure}}
\setcounter{figure}{0}
\renewcommand{\thetable}{C.\arabic{table}}
\setcounter{table}{0}
\renewcommand{\theequation}{C.\arabic{equation}}
\setcounter{equation}{0}
\renewcommand{\thefootnote}{C.\arabic{footnote}}
\setcounter{footnote}{0}

\section{Additional equations for the consumer and planner problems}
\subsection{Great Recession}
\label{sec:Appendix_value_function_details-2008}
Let $V^{U}\left(\omega\right)$ denote the value function for an agent of type $\omega$ that is unemployed for at least part of the second crisis-period, $\xi<1$. Because the shock is transitory, the economy will transition back to the pre-crisis equilibrium, where the value function is as given in Equation~\eqref{eq:Value-function}. We get the following expression for $V^{U}\left(\omega\right)$:\footnote{See Appendix Section~\ref{sec:Appendix_Calibration} for details on how spousal income is affected by unemployment.}
\begin{equation}\begin{array}{lll}
V^{U}\left(\omega\right) & = & \underset{c\geq0,a^{\prime}\geq0}{\max}\,\,\,u\left(c,m,k\right)+\delta\psi_{j}\text{\ensuremath{\mathbb{E}}}_{\eta^{\prime}\vert\eta}\text{\ensuremath{\mathbb{E}}}_{\nu^{\prime}\vert\nu}\text{\ensuremath{\mathbb{E}}}_{k^{\prime}\vert\left(j,e,m,k\right)}V\left(\omega^{\prime}\right)\\
 & \text{s.t.} & c+a^{\prime}=a+y-T_y\\
 &  & y=ra+\mathbbm{1}_{j<j_{R}}\theta\epsilon_{j,\eta,e}\left[\xi+b\left(1-\xi\right)\right]+\mathbbm{1}_{j\geq j_{R}}SS_{e}+\mathbbm{1}_{m=1}B_{j,e,k,\eta,\nu,\xi,b}.
\end{array}\label{eq:V_U-Value-function-2008}
\end{equation}

\subsection{COVID-19}\label{sec:Appendix_value_function_details-2021}
Let $\tilde{x}\left(y,j,m,k;D\right)$ denote the ex ante expected welfare in 2021 of agents with income $y$, age $j$, marital status $m$, and number of children $k$ in 2020 that receive $D$:
\begin{equation}
\begin{array}{lll}
\tilde{x}\left(y,j,m,k;D\right) & = & \psi_{j}\text{\ensuremath{\mathbb{E}}}_{\eta^{\prime}\vert\eta}\text{\ensuremath{\mathbb{E}}}_{\nu^{\prime}\vert\nu}\text{\ensuremath{\mathbb{E}}}_{k^{\prime}\vert\left(j,e,m,k\right)}\ensuremath{\mathbb{E}}_{\delta\vert e}\ensuremath{\mathbb{E}}_{(a,\eta,e,\nu)\vert(y,j,m,k)}\\
 &  & \Big[\pi^{U}\left(j+1,\eta^{\prime},e\right)\overline{x}^{U}\left(j+1,a'\left(\omega\right),\eta^{\prime},e,m,k^{\prime},\delta,\nu^{\prime};D\right)\\
 &  & +\left(1-\pi^{U}\left(j+1,\eta^{\prime},e\right)\right)\overline{x}^{W}\left(j+1,a' \left(\omega\right),\eta^{\prime},e,m,k^{\prime},\delta,\nu^{\prime};D\right)\Big],
\end{array}\label{eq:x_tilde}
\end{equation}
where $a'\left(\omega\right)$, $\ensuremath{\mathbb{E}}_{\delta\vert e}$, and $\ensuremath{\mathbb{E}}_{(a,\eta,e,\nu)\vert(y,j,m,k)}$ are as defined in Section~\ref{sec:Planner-problem-2008} and $\overline{x}^{q}\left(\cdot;D\right)$ ($q\in\{U,W\}$) is as defined in Equation \eqref{eq:constant-cons-stream-V_q}.
\clearpage 

\renewcommand{\thefigure}{D.\arabic{figure}}
\setcounter{figure}{0}
\renewcommand{\thetable}{D.\arabic{table}}
\setcounter{table}{0}
\renewcommand{\theequation}{D.\arabic{equation}}
\setcounter{equation}{0}
\renewcommand{\thefootnote}{D.\arabic{footnote}}
\setcounter{footnote}{0}

\section{Technical details}
\label{sec:Appendix_computational_details}
\textit{Proof that the welfare measure preserves the order of the value functions}---As discussed in Section~\ref{sec:Planner-problem-2021}, we apply the following conversion when calculating welfare:
\begin{align}
\overline{x}^{q}(\omega;D)=x^{q}(\omega;D)\Lambda(\omega)^{\frac{1}{1-\gamma}}=\left[V^{q}(\omega;D)(1-\gamma)\right]^{\frac{1}{1-\gamma}},\,\,\,\,q\in\{U,W\}.
\end{align}
where $x^{q}(\omega;D)$ is as given in Equation \eqref{eq:constant-cons-stream-V_q}. This section verifies that our conversion preserves the ordinal rank of $V^{q}(\omega;D)$ across types $\omega$ for all $\gamma$.

\begin{proof}
Consider two agents with idiosyncratic type $\omega_1$ and $\omega_2$. Suppose $V^{q}(\omega_1;D)>V^{q}(\omega_2;D)$. Applying our conversion:
\begin{equation}\label{eq:welfare_measure_proof}
\begin{array}{ccc}
\overline{x}^{q}(\omega_1;D)=x^{q}(\omega_1;D)\Lambda(\omega_1)^{\frac{1}{1-\gamma}} & = & \left[V^{q}(\omega_1;D)(1-\gamma)\right]^{\frac{1}{1-\gamma}} \\
\overline{x}^{q}(\omega_2;D)=x^{q}(\omega_2;D)\Lambda(\omega_1)^{\frac{1}{1-\gamma}} & = & \left[V^{q}(\omega_2;D)(1-\gamma)\right]^{\frac{1}{1-\gamma}}
\end{array}
\end{equation}
We want to show that $V^{q}(\omega_1;D)>V^{q}(\omega_2;D)$ implies that $\overline{x}^{q}(\omega_1;D)>\overline{x}^{q}(\omega_2;D)$. This simplifies to showing that $\left[V^{q}(\omega_1;D)(1-\gamma)\right]^{\frac{1}{1-\gamma}}>\left[V^{q}(\omega_2;D)(1-\gamma)\right]^{\frac{1}{1-\gamma}}$ for all $\gamma$. If $\gamma\in[0,1)$, the inequality can be rewritten as $V^{q}(\omega_1;D)(1-\gamma)>V^{q}(\omega_2;D)(1-\gamma)$, which holds by assumption (the case where $\gamma=1$ follows trivially). If $\gamma>1$, the inequality can be rewritten as $V^{q}(\omega_1;D)(1-\gamma)<V^{q}(\omega_2;D)(1-\gamma)$, which can further be rewritten as $V^{q}(\omega_1;D)>V^{q}(\omega_2;D)$.
\end{proof}

\textit{Computational details}---For each household demographic type (marital status, number of children, and age of household head), the allocation problems require us to compute the value of receiving a particular tax rebate or check amount at very small household income intervals because the amounts under the actual policies varied with $\$$5 increments in household income. This requires a dense state space. We have 83 age groups, 5 child states, 2 marital states, 2 educational states, 2 discount factor states, 65 asset states, and 1,330 labor productivity states, resulting in an idiosyncratic state space of 287,014,000 elements. The model is solved with a continuous choice for saving through backwards induction by means of the vectorized bisection algorithm described in \textcite{wangEmpiricalEquilibriumModel2021a}.

To derive the optimal allocations, we need to compute each household's value of receiving a particular tax rebate or check amount. Because a large share of the households will be unemployed during the crisis periods, we need to compute these values conditional on employment status (employed or unemployed). For the 2020--2021 crisis, we first discretize the check from $\$$0--$\$$16,800 in $\$$100 dollar increments. For each of the 287,014,000 idiosyncratic types, we then compute the value of receiving a particular amount $D\in\left\{0,100,\dots,16800\right\}$ conditional on employment status. Because households have the option of saving all or part of their checks, this requires us to solve for $287,014,000\times169\times2=97,010,732,000$ different household-check values. We follow an analogous approach for the 2008--2009 crisis.

Consistent with the actual policies, eligibility for the tax rebates and checks in the model is tied to each household's income and family size (and hence tax-liability) in the previous period. For the 2020--2021 crisis, we split households into 399,230 groups, where groups are defined by marital status, number of children, age of household head, and household income. We have 450 household income groups in $\$$500 increments from $\$$0--$\$$225,000, and 31 income groups in $\$$5,000 increments from $\$$225,000--$\$$383,000. Given our discretization of the check amount, this leads to $399,230\times(169-1)=67,070,640$ different marginal values of checks. Our approach for the 2008--2009 crisis is analogous.

\textit{Definition of equilibrium}---Given a fiscal policy $\left\{SS_{e=0}, SS_{e=1}, T_y,G\right\}$ and a real interest rate $r$, the pre-crisis steady-state competitive equilibrium consists of household policies $\left\{c(\omega),a^\prime(\omega)\right\}$ and an associated value function $\left\{V(\omega)\right\}$ such that:
\begin{enumerate}
  \item Given prices and fiscal policy, consumers maximize utility subject to their constraints.
  \item Government policies balance the government budget constraint: $\int T_y(\omega)\mu(d\omega)=G+\int SS_{e}\mu(d\omega)$, where $y=ra+\mathbbm{1}_{j<j_{R}}\theta\epsilon_{j,\eta,e}+\mathbbm{1}_{j\geq j_{R}}SS_e+\mathbbm{1}_{m=1}B_{j,e,k,\eta,\nu}$ is household income.\footnote{See the \href{https://fanwangecon.github.io/PrjOptiSNW/}{companion website} for the model whereby, in the event of death, the consumer's accidental bequests are transferred to the government.}
  \item The measure of agents of type $\omega=(j,a,\eta,e,m,k,\delta,\nu)$, $\mu(\omega)$, is induced by the exogenous initial distribution, the policy functions, the age-specific mortality risk, and the exogenous stochastic processes for idiosyncratic shocks.
\end{enumerate}
\clearpage 

\renewcommand{\thefigure}{E.\arabic{figure}}
\setcounter{figure}{0}
\renewcommand{\thetable}{E.\arabic{table}}
\setcounter{table}{0}
\renewcommand{\theequation}{E.\arabic{equation}}
\setcounter{equation}{0}
\renewcommand{\thefootnote}{E.\arabic{footnote}}
\setcounter{footnote}{0}

\section{Outcomes as functions of the optimal allocation queue}
\label{appendix:sec:Additional_formulas}
Let ${W}$ represent some level of resources. Let $\left\{D_g^o\right\}_{g=1}^{N}$ be a vector of resource exhausting allocations given $W$, such that $W=\sum_{g=1}^{N} D_g^o$. Let $\Delta^{W}>0$ be some increment in total resources available, and let $Q_{g,l}$ denote the position of the $l$'th increment to group $g$ in an optimal allocation queue as defined in Equation \eqref{eq:optimal_allocation_queue}.

\subsection{Comparing optimal to alternative policy for given budget}

The increase in aggregate outcome, $\sum_{g=1}^N H_g \left(D_g\right)$ (aggregate consumption for example), achieved by the optimal versus an alternative set of resource exhausting allocations is:
\begin{equation}
\sum_{g=1}^{N}
\left(
\sum_{l=1}^{\bar{D}_g}
\left(
    \alpha_{g,l}
    \cdot
    \mathbbm{1}
    \left\{
        Q_{g,l} \le W
    \right\}
\right)
-
\sum_{l=1}^{D_g^{o}}
    \alpha_{g,l}
\right).
\end{equation}

\subsection{Impact of additional resources on the optimal aggregate outcomes}

When a planner allocates along the optimal allocation queue, the increase in aggregate outcome achieved with an increase in aggregate resources is given by:
\begin{align}
\sum_{g=1}^{N}
\sum_{l=1}^{\bar{D}_g}
\left(
    \alpha_{g,l}
    \cdot
    \mathbbm{1}
    \left\{
        {W}
        <
        Q_{g,l}
        \le
        \left({W} + \Delta^{W}\right)
    \right\}
\right).
\end{align}

\subsection{Elasticity of optimal aggregate outcome with respect to resources}

Along the optimal allocation queue, the elasticity of the aggregate outcome with respect to resource increment $\Delta^{W}$ at current resources level $W$ is given by:
\begin{align}
\epsilon
\left(
{W},
\Delta^{W}
\right)
=
\frac{\sum_{g=1}^{N}
\sum_{l=1}^{\bar{D}_g}\left(
    \alpha_{g,l}
    \cdot
    \mathbbm{1}
    \left\{
        {W} \le Q_{g,l} \le \left({W} + \Delta^{W} \right)
    \right\}
\right)}{\sum_{g=1}^{N}
\left(
A_g +
\sum_{l=1}^{\bar{D}_g}
\left(
    \alpha_{g,l}
    \cdot
    \mathbbm{1}
    \left\{
        Q_{g,l} \le {W}
    \right\}
\right)
\right)}
\cdot
\frac{{W}}{\Delta^{W}}.
\end{align}
When the outcome is consumption, this is the maximum---among all constrained allocation alternatives---elasticity of aggregate consumption with respect to the an increase in total stimulus.

\subsection{Gini-index as a function of the optimal allocation queue}
An individual outcome given aggregate resources and the optimal allocation queue can be expressed as:
\begin{align}
H_g\left({W}, \mathbf{Q}_g\right)
=
A_g +
\sum_{l=1}^{\bar{D}_g}
\left(
    \alpha_{g,l}
    \cdot
    \mathbbm{1}
    \left\{
        Q_{g,l} \le {W}
    \right\}
\right) \, ,
\end{align}
where $\mathbf{Q}_g$ is the vector of queue positions for allocations to group $g$. The Gini coefficient is a function of overall resources ${W}$ and the full queue $\mathbf{Q}$ for all allocation positions:\\
\\
\resizebox{1.0\hsize}{!}{
$\text{GINI}\left({W}, \mathbf{Q}\right)
=
1
-
\left(\frac{2}{N+1}\right)
\cdot
\left(
\frac{
\sum_{j=1}^{N}
\sum_{g=1}^{N}
\left(
H_g\left({W}, \mathbf{Q}_g\right)
\cdot
\mathbbm{1}
\left(
H_g\left({W}, \mathbf{Q}_g\right)
\le
H_j\left({W}, \mathbf{Q}_j\right)
\right)
\right)
}
{
\sum_{g=1}^{N}
H_g\left({W}, \mathbf{Q}_g\right)
}
\right)$}
\clearpage

\end{document}